\documentclass[twocolumn]{aastex631}

\usepackage{amsmath}
\usepackage{enumitem}
\setlist{nolistsep}

\shorttitle{Inclination-dependent UV SFRs}
\shortauthors{Doore et al.}

\begin{document}

\title{The Impact of Inclination-dependent Attenuation on Ultraviolet Star Formation Rate Tracers}

\correspondingauthor{Keith Doore}
\email{kjdoore@uark.edu}

\author[0000-0001-5035-4016]{Keith~Doore}
\affiliation{Department of Physics, University of Arkansas, 226 Physics Building, 825 West Dickson Street, Fayetteville, AR 72701, USA}

\author[0000-0002-2987-1796]{Rafael~T.~Eufrasio}
\affiliation{Department of Physics, University of Arkansas, 226 Physics Building, 825 West Dickson Street, Fayetteville, AR 72701, USA}

\author[0000-0003-2192-3296]{Bret~D.~Lehmer}
\affiliation{Department of Physics, University of Arkansas, 226 Physics Building, 825 West Dickson Street, Fayetteville, AR 72701, USA}

\author[0000-0001-8473-5140]{Erik~B.~Monson}
\affiliation{Department of Physics, University of Arkansas, 226 Physics Building, 825 West Dickson Street, Fayetteville, AR 72701, USA}

\author[0000-0001-8525-4920]{Antara~Basu-Zych}
\affiliation{NASA Goddard Space Flight Center, Code 662, Greenbelt, MD 20771, USA}
\affiliation{Center for Space Science and Technology, University of Maryland Baltimore County, 1000 Hilltop Circle, Baltimore, MD 21250, USA}

\author[0000-0002-9202-8689]{Kristen~Garofali}
\affiliation{NASA Goddard Space Flight Center, Code 662, Greenbelt, MD 20771, USA}



\begin{abstract}

We examine and quantify how hybrid (e.g., UV+IR) star formation rate (SFR) estimators and the $A_{\rm FUV}$--$\beta$ relation depend on inclination for disk-dominated galaxies using spectral energy distribution modeling that utilizes the inclination-dependent attenuation curves described in Doore et al. We perform this analysis on a sample of 133 disk-dominated galaxies from the CANDELS fields and 18 disk galaxies from the Spitzer Infrared Nearby Galaxies Survey and Key Insights on Nearby Galaxies: A Far-Infrared Survey with Herschel samples. We find that both the hybrid SFR estimators and the $A_{\rm FUV}$--$\beta$ relation present clear dependencies on inclination. To quantify this dependence in the hybrid SFR estimators, we derive an inclination and a far-UV–-near-IR color-dependent parametric relation for converting observed UV and IR luminosities into SFRs. For the $A_{\rm FUV}$--$\beta$ relation, we introduce an inclination-dependent component that accounts for the majority of the inclination dependence with the scatter of the relation increasing with inclination. We then compare both of these inclination-dependent relations to similar inclination-independent relations found in the literature. From this comparison, we find that the UV+IR correction factor and $A_{\rm FUV}$ for our hybrid and $A_{\rm FUV}$--$\beta$ relations, respectively, result in a reduction in the residual scatter of our sample by approximately a factor of 2. Therefore, we demonstrate that inclination must be considered in hybrid SFR estimators and the $A_{\rm FUV}$--$\beta$ relation to produce more accurate SFR estimates in disk-dominated galaxies.

\end{abstract}



\section{Introduction} \label{sec:Intro}

Stars are one of the basic building blocks of galaxies, and measurements of their formation rates are critical for understanding how galaxies assembled and evolved. On extragalactic scales,  star formation rates (SFRs) are typically determined for subgalactic star forming regions \citep[e.g.,][]{2008AJ....136.2846B,2012AJ....144....3L,2014ApJ...795...89E,2017ApJ...851...10E,2019MNRAS.482L..55T} or, more commonly, entire integrated galaxies \citep[e.g.,][]{1983ApJ...272...54K,2004ApJ...606..271G,2007ApJS..173..267S,2013A&A...558A..67A,2019ApJS..243...22B}. At these scales, SFRs are typically determined from basic parametric descriptions (e.g. hybrid estimators, \citealt{1999ApJ...521...64M} relation, etc.), rather than physically based characterizations of the galaxy or each star forming region \citep[see][for a review]{2012ARA&A..50..531K}. Therefore, to improve estimates of SFRs, these parametric descriptions can be expanded to include dependencies on physical properties relevant to the SFR calculation.

Generally, parameterizations of SFRs use intrinsic (i.e., unattenuated) ultraviolet (UV) emission, which is almost exclusively produced by emission from young ($\leqslant$a few hundred Myr), massive stars,
\begin{equation}
\bigg(\frac{\rm SFR}{M_\odot\ \rm{yr}^{-1}}\bigg)=k_{\rm UV}\bigg(\frac{L^{\rm intr}_{\rm UV}}{L_\odot}\bigg),
\label{eq:LUVtoSFR}
\end{equation}
where $k_{\rm UV}$ is the conversion from the intrinsic monochromatic luminosity in the UV ($L^{\rm intr}_{\rm UV}$, calculated as $\nu L_\nu$) to the average SFR over the past 100 Myr \citep{1998ARA&A..36..189K,2011ApJ...737...67M,2012ARA&A..50..531K}. The conversion factor $k_{\rm UV}$ is typically determined from stellar population synthesis and depends upon the chosen UV bandpass filter, initial mass function (IMF), metallicity, and assumed star formation history (SFH, the SFR as a function of time).

Unlike $k_{\rm UV}$, which can be determined theoretically with basic assumptions, $L^{\rm intr}_{\rm UV}$ is more difficult to determine, since the true intrinsic luminosity cannot be measured directly due to attenuation by dust. Instead, $L^{\rm intr}_{\rm UV}$ must be estimated by modeling the attenuation of the observed emission in the rest-frame UV. There are two common methods for doing this, depending on the availability of quality infrared (IR) data. If quality IR data are available, hybrid SFR estimators are often chosen \citep[e.g.,][]{2008AJ....136.2782L,2008ApJ...686..155Z,2011ApJ...741..124H,2014ApJ...795...89E,2015A&A...584A..87C,2016A&A...591A...6B,2017ApJ...851...10E}. These tracers correct the observed UV luminosity to an intrinsic UV luminosity by assuming that some fraction of the attenuated UV light is absorbed by dust and reradiated in the IR, or
\begin{equation}
L^{\rm intr}_{\rm UV}=L^{\rm obs}_{\rm UV}+a_{\rm corr} \times L^{\rm obs}_{\rm IR},
\label{eq:acorr}
\end{equation}
where $L^{\rm obs}_{\rm UV}$ is the observed rest-frame UV luminosity assuming isotropy, $a_{\rm corr}$ is the UV+IR correction factor that accounts for some fraction of the reradiated IR emission being from the attenuated UV light, and $L^{\rm obs}_{\rm IR}$ is the observed emission in a rest-frame IR bandpass, or the total integrated IR (TIR) luminosity. Many values of $a_{\rm corr}$ exist in the literature that have been empirically derived depending upon the chosen UV and IR bandpasses, as well as the choice of attenuation curve.

Another commonly used method for modeling the attenuation of the UV emission when IR data are not available is the $A_{\rm UV}$--$\beta$ relation, which is also referred to as the \citet{1999ApJ...521...64M} relation due to its initial derivation in \citet{1999ApJ...521...64M}. This relation links the slope of the observed UV emission ($\beta$; $F_\lambda \varpropto \lambda^{\beta}$) to the UV attenuation ($A_{\rm UV}$). Following the notation of \citet{2012A&A...539A.145B}, a generalized version of the $A_{\rm UV}$--$\beta$ relation is given by
\begin{equation}
A_{\rm UV}=a_\beta(\beta-\beta_0),
\label{eq:AUVbeta}
\end{equation}
where $\beta_0$ is the slope of the unattenuated UV emission given by the galaxy's intrinsic properties (i.e., SFH, IMF, and metallicity), and $a_\beta$ is defined by the shape of the chosen attenuation curve. This relation is commonly calibrated using a sample of galaxies that have IR measurements to use their ``IR excess'' (IRX) as a proxy for $A_{\rm UV}$ \citep{1994ApJ...429..582C,1999ApJ...521...64M,2000ApJ...533..236G,2004MNRAS.349..769K,2011ApJ...741..124H,2012A&A...539A.145B,2012A&A...545A.141B}. This leads to the so-called IRX--$\beta$ relation, given by
\begin{equation}
{\rm IRX} \equiv \log_{10} \Bigg(\frac{L_{\rm IR}^{\rm obs}}{L_{\rm UV}^{\rm obs}}\Bigg) = \\
\log_{10}\Big[\Big(10^{0.4a_\beta(\beta-\beta_0)}-1\Big)/a_{\rm corr}\Big],
\end{equation}
where $a_{\rm corr}$ is defined in Equation~\ref{eq:acorr}. Once $a_\beta$, $\beta_0$, and $a_{\rm corr}$ have been calibrated, the $A_{\rm UV}$--$\beta$ relation can be used to determine the deattenuated, intrinsic UV luminosity for galaxies lacking IR data.

However, both of these methods have a common caveat. As stated above, the parameters $a_{\rm corr}$ and $a_\beta$ strongly depend upon the choice of attenuation curve. Therefore, a simplified or inappropriate choice of attenuation curve can lead to various biases in these values. This is of particular importance when trying to determine the intrinsic UV emission of disk galaxies, as the inclination of the disk has been shown to significantly influence attenuation, with edge-on galaxies (i.e., $i\approx90^\circ$) having increased attenuation compared to face-on galaxies \citep[i.e., $i\approx0^\circ$;][]{1994AJ....107.2036G,2007MNRAS.379.1022D,2008ApJ...687..976U,2010ApJ...718..184C,2010MNRAS.404..792M,2011MNRAS.417.1760W,2016MNRAS.459.2054D,2017ApJ...851...90B,2018ApJ...859...11S}.

As an example, if a disk galaxy could be viewed from multiple inclinations, it would be observed that the UV emission would decrease with increasing inclination, whereas the IR emission would be relatively unchanged due to minimal attenuation at these wavelengths. With the intrinsic UV emission being independent of inclination, Equation~\ref{eq:acorr} indicates that $a_{\rm corr}$ must be dependent upon inclination to compensate for the inclination dependence of the observed UV emission. Therefore, in order to account for this effect and obtain accurate SFR estimators, it is critical to characterize how inclination affects the attenuation and scaling relations of disk galaxies.

Recent works by \citet{2010ApJ...718..184C}, \citet{2018A&A...615A...7L,2018A&A...616A.157L}, \citet{2018ApJ...869..161W}, and \citet{2018MNRAS.480.3788W} have investigated how inclination affects the SFRs derived using UV emission. Specifically, \citet{2018A&A...615A...7L,2018A&A...616A.157L} and \citet{2018MNRAS.480.3788W} showed that inclination-based attenuation alone can cause the uncorrected, observed UV emission to yield underestimated SFRs (by factors of 2.5--4) for edge-on galaxies compared to face-on galaxies. \citet{2010ApJ...718..184C} and \citet{2018ApJ...869..161W} showed that the IRX--$\beta$ relation is highly dependent upon inclination, with nearly edge-on galaxies having larger IRX values by factors of 1.2--1.5 compared to nearly face-on galaxies with the same $\beta$. However, \citet{2018A&A...616A.157L} showed that hybrid SFR estimators, when assuming a constant $a_{\rm corr}$, are relatively inclination-independent when compared to the galaxy main sequence (galaxy SFR--stellar mass relation). Yet this is not in contradiction with the theoretical stance that hybrid SFR estimators, when assuming a constant $a_{\rm corr}$, should be dependent upon inclination. This is due to the comparison with the galaxy main sequence, which was derived using these same hybrid SFR estimators. Therefore, it is expected that any trends with inclination are masked by using this comparison.

In this paper, we examine and quantify how both hybrid SFR estimators and the $A_{\rm UV}$--$\beta$ relation depend on inclination using spectral energy distribution (SED) modeling that incorporates the inclination-dependent attenuation curves described in \citet{2021ApJ...923...26D}, which are based on the \citet{2004A&A...419..821T} inclination-dependent attenuation curves. When examining this dependence, we specifically focus on the commonly used Galaxy Evolution Explorer (GALEX) far-UV (FUV) bandpass and TIR luminosity ($L_{\rm TIR}$). We quantify this inclination dependence using a sample of 133 galaxies from the Cosmic Assembly Near-infrared Deep Extragalactic Legacy Survey (CANDELS) fields \citep{2011ApJS..197...35G,2011ApJS..197...36K} along with 18 disk galaxies from the Spitzer Infrared Nearby Galaxies Survey \citep[SINGS;][]{2003PASP..115..928K,2005ApJ...633..857D,2007ApJ...655..863D} and Key Insights on Nearby Galaxies: A Far-Infrared Survey with Herschel \citep[KINGFISH;][]{2011PASP..123.1347K,2012ApJ...745...95D} samples. We discuss how we selected these galaxies and their photometry in Section~\ref{sec:Data}. In Section~\ref{sec:Derive}, we derive the physical properties needed for our analysis using SED modeling. In Section~\ref{sec:AnalysisDiscuss}, we examine, quantify, and present how both the hybrid SFR estimators and the $A_{\rm FUV}$--$\beta$ relation depend on inclination and discuss how this inclination dependence compares with results from past studies. Finally, we summarize our results in Section~\ref{sec:Summary}.

In this work, we assume a \citet{2001MNRAS.322..231K} IMF with solar metallicity ($Z=Z_\odot$) and a flat $\Lambda$CDM cosmology where $\Omega_M=0.30$ and $\Omega_{\Lambda}=0.70$ with a Hubble constant of $H_0=70~\rm{km~s}^{-1}~\rm{Mpc}^{-1}$. Additionally, all quoted magnitudes are in AB magnitudes.

\section{Data and Sample Selection} \label{sec:Data}

\subsection{CANDELS sample} \label{sec:CANDELS}

Since UV star formation tracers are commonly used to determine the SFRs of galaxies at intermediate redshifts, we utilized a sample of 133 disk-dominated galaxies that are contained within the CANDELS fields, spanning a redshift range of $z=0.09$--0.98. Of these galaxies, 38 and 42 galaxies are contained within the Great Observatories Origins Deep Survey North (GOODS-N) and South (GOODS-S) fields \citep{2004ApJ...600L..93G}, respectively; 23 are contained within the Extended Groth Strip \citep[EGS;][]{2007ApJ...660L...1D}; 25 are contained within the Cosmic Evolution Survey (COSMOS) field \citep{2007ApJS..172....1S}; and five are contained within the UKIDSS Ultra-Deep Survey (UDS) field \citep{2007MNRAS.380..585C,2007MNRAS.379.1599L}. To generate this sample of galaxies, we used a similar selection method as presented in \citet{2021ApJ...923...26D}, which was shown to have minimal to no selection biases due to inclination. 

We briefly summarize this method here. We first selected galaxies to have reliable spectroscopic redshifts from our compiled spectroscopic redshift catalog, which is described in Appendix~\ref{sec:RedCatalog}. We then required each galaxy to have at least six photometric measurements in the mid-to-far IR (3--1000~$\mu$m), one of which was required to be greater than 100~$\mu$m rest frame to constrain the peak of the dust emission. Next, we considered any galaxy cross-matched within $1^{\prime\prime}$ of an X-ray-detected source in the Chandra X-ray catalogs \citep{2015ApJS..220...10N,2016ApJ...819...62C,2016ApJS..224...15X,2017ApJS..228....2L,2018ApJS..236...48K} as potentially harboring an active galactic nucleus (AGN). These potential AGNs were then removed to prevent any AGN-dominated galaxies from being in the sample. We also removed potentially obscured mid-IR AGNs using the \citet{2012ApJ...748..142D} IRAC selection criteria and \citet{2013ApJ...763..123K} Spitzer/Herschel color-color criteria. We then reduced the sample to only disk-dominated galaxies (i.e., an approximate bulge-to-disk ratio of zero) via their S\'ersic index $n$ \citep[$n<1.2$;][]{1963BAAA....6...41S} as measured by \citet{2012ApJS..203...24V}\footnote{\url{https://users.ugent.be/~avdrwel/research.html\#candels}} in the Hubble Space Telescope (HST) WFC3/F125W band. We additionally required the S\'ersic indices to be from ``good fits'' (i.e., flag of 0). Finally, a visual inspection of HST postage stamps was performed, and we removed any irregular or potentially merging galaxies that survived the S\'ersic index cut.

\begin{figure}[t!]
\centerline{
\includegraphics[width=8.75cm]{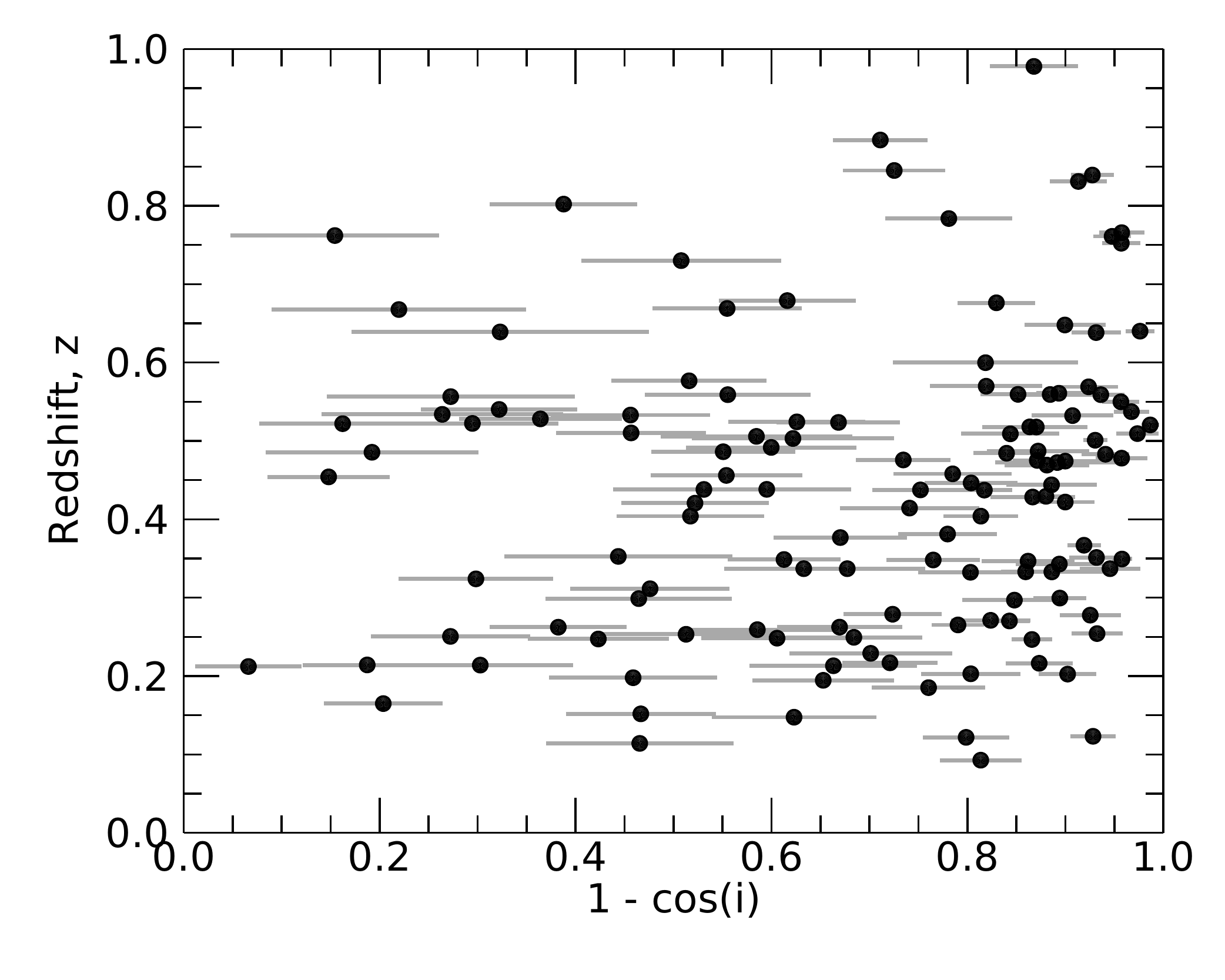}
}
\caption{
Inclinations derived from \texttt{Lightning} in terms of $1-\cos i$ vs. the spectroscopic redshift of each galaxy in the CANDELS sample. While the sample does contain more inclined galaxies compared to less inclined galaxies, there is no distinguishable trend in inclination with redshift.
}
\label{fig:incvred}
\end{figure}

To confirm that minimal to no selection biases due to inclination are present in our sample, we show the inclination of each galaxy as derived from our SED fittings (see Section~\ref{sec:Lightning}) versus spectroscopic redshift in Figure~\ref{fig:incvred}. While there are more highly inclined galaxies compared to low-inclination galaxies, no distinguishable trend in inclination with redshift is present. Trends between inclination and redshift are possible, as edge-on galaxies can be preferentially selected at higher redshifts compared to face-on galaxies due to their higher surface brightness \citep{2008MNRAS.388.1708G,2010ApJ...714L.113S,2016MNRAS.459.2054D}. We quantitatively confirmed this lack of trend between inclination and redshift by splitting the sample into two groups along the median redshift of 0.45 and performing a Kolmogorov--Smirnov test. The test showed minimal differences in inclination distributions for the high- and low-redshift groups with a $p$-value $>0.1$.

The UV--to--mid-IR photometry for the 133 galaxies was taken from the CANDELS multiband photometric catalogs\footnote{\url{https://archive.stsci.edu/prepds/candels/}}, which are presented in \citet{2019ApJS..243...22B}, \citet{2013ApJS..207...24G}, \citet{2017ApJS..229...32S}, \citet{2017ApJS..228....7N},  and \citet{2013ApJS..206...10G} for the GOODS-N, GOODS-S, EGS, COSMOS, and UDS fields, respectively. We also utilized the far-IR photometry produced by \citet{2019ApJS..243...22B} for all five of the CANDELS fields. We corrected the photometry for Galactic extinction using the \citet{2011ApJ...737..103S} recalibration of the \citet{1998ApJ...500..525S} dust maps and a \citet{1999PASP..111...63F} reddening law with $R_V = 3.1$. The extinction was determined for the center of each field, and no variation across each field is considered, due to small overall extinction corrections and minimal variation across each field. We also added fractional calibration uncertainties to the catalog flux uncertainties to account for any additional sources of uncertainty and potential systematic variations in the photometry. These fractional calibration uncertainties are 2--15\% of the measured flux as described in each instrument's user handbook and listed in Table~\ref{table:CANDELSSED} along with the mean wavelength, Galactic extinction, and corresponding filters used in each field.

To estimate the inclinations of each galaxy (see Section~\ref{sec:Lightning}), we required an axis ratio~$q$ with uncertainty. Therefore, we utilized the WFC3/F125W measured axis ratios from the fits for the S\'ersic index by \citet{2012ApJS..203...24V}. We note that measurements of $q$ have been shown to vary with rest-frame wavelength and redshift \citep{2002AJ....124.1328D}. However, \citet{2014ApJ...792L...6V} showed that this variation with redshift in the \citet{2012ApJS..203...24V} axis ratios is generally smaller than the uncertainty within our redshift range.

\startlongtable
\begin{deluxetable*}{llcccllccc}
\tabletypesize{\footnotesize}
\tablecaption{CANDELS Multiwavelength Coverage \label{table:CANDELSSED}}
\tablecolumns{10}
\tablehead{ Field & Telescope/Band & $\lambda_{\rm mean}$\tablenotemark{a} & $A_{\lambda}^{\rm Gal}$\tablenotemark{b} & $\sigma_{\rm C}^{\rm cal}$\tablenotemark{c}  & Field & Telescope/Band & $\lambda_{\rm mean}$\tablenotemark{a} & $A_{\lambda}^{\rm Gal}$\tablenotemark{b} & $\sigma_{\rm C}^{\rm cal}$\tablenotemark{c}  \\ & & ($\mu$m) & (mag) &  & & & ($\mu$m) & (mag) & }
\startdata
GOODS-N  & KPNO 4m/Mosaic $U$           & 0.3561   & 0.052  & 0.05 & EGS      & CFHT/MegaCam $u^{*}$         & 0.3799   & 0.032  & 0.05 \\
         & LBT/LBC $U$                  & 0.3576   & 0.052  & 0.10 &          & CFHT/MegaCam $g^{\prime}$    & 0.4806   & 0.026  & 0.05 \\
         & HST/ACS F435W                & 0.4296   & 0.044  & 0.02 &          & HST/ACS F606W                & 0.5804   & 0.020  & 0.02 \\
         & HST/ACS F606W                & 0.5804   & 0.031  & 0.02 &          & CFHT/MegaCam $r^{\prime}$    & 0.6189   & 0.018  & 0.05 \\
         & HST/ACS F775W                & 0.7656   & 0.020  & 0.02 &          & CFHT/MegaCam $i^{\prime}$    & 0.7571   & 0.013  & 0.05 \\
         & HST/ACS F814W                & 0.7979   & 0.019  & 0.02 &          & HST/ACS F814W                & 0.7979   & 0.012  & 0.02 \\
         & HST/ACS F850LP               & 0.8990   & 0.015  & 0.02 &          & CFHT/MegaCam $z^{\prime}$    & 0.8782   & 0.011  & 0.05 \\
         & HST/WFC3 F105W               & 1.0449   & 0.012  & 0.02 &          & Mayall/NEWFIRM $J_1$         & 1.0432   & 0.008  & 0.10 \\
         & HST/WFC3 F125W               & 1.2396   & 0.009  & 0.02 &          & Mayall/NEWFIRM $J_2$         & 1.1922   & 0.006  & 0.10 \\
         & HST/WFC3 F140W               & 1.3784   & 0.007  & 0.02 &          & HST/WFC3 F125W               & 1.2396   & 0.006  & 0.02 \\
         & HST/WFC3 F160W               & 1.5302   & 0.006  & 0.02 &          & CFHT/WIRCam $J$              & 1.2513   & 0.006  & 0.05 \\
         & CFHT/WIRCam $K_s$            & 2.1413   & 0.004  & 0.05 &          & Mayall/NEWFIRM $J_3$         & 1.2757   & 0.006  & 0.10 \\
         & Subaru/MOIRCS $K_s$          & 2.1442   & 0.004  & 0.05 &          & HST/WFC3 F140W               & 1.3784   & 0.005  & 0.02 \\
         & Spitzer/IRAC1                & 3.5314   & 0.002  & 0.05 &          & HST/WFC3 F160W               & 1.5302   & 0.004  & 0.02 \\
         & Spitzer/IRAC2                & 4.4690   & 0.000  & 0.05 &          & Mayall/NEWFIRM $H_1$         & 1.5578   & 0.004  & 0.10 \\
         & Spitzer/IRAC3                & 5.6820   & 0.000  & 0.05 &          & CFHT/WIRCam $H$              & 1.6217   & 0.004  & 0.05 \\
         & Spitzer/IRAC4                & 7.7546   & 0.000  & 0.05 &          & Mayall/NEWFIRM $H_2$         & 1.7041   & 0.004  & 0.10 \\
         & Spitzer/MIPS 24 $\mu$m       & 23.513   & 0.000  & 0.05 &          & CFHT/WIRCam $K_s$            & 2.1413   & 0.002  & 0.05 \\
         & Spitzer/MIPS 70 $\mu$m       & 70.389   & 0.000  & 0.10 &          & Mayall/NEWFIRM $K$           & 2.1639   & 0.002  & 0.10 \\
         & Herschel/PACS 100 $\mu$m     & 100.05   & 0.000  & 0.05 &          & Spitzer/IRAC1                & 3.5314   & 0.001  & 0.05 \\
         & Herschel/PACS 160 $\mu$m     & 159.31   & 0.000  & 0.05 &          & Spitzer/IRAC2                & 4.4690   & 0.000  & 0.05 \\
         & Herschel/SPIRE 250 $\mu$m    & 247.21   & 0.000  & 0.15 &          & Spitzer/IRAC3                & 5.6820   & 0.000  & 0.05 \\
\cline{1-5}
GOODS-S  & Blanco/MOSAIC II $U$         & 0.3567   & 0.034  & 0.05 &          & Spitzer/IRAC4                & 7.7546   & 0.000  & 0.05 \\
         & VLT/VIMOS $U$                & 0.3709   & 0.033  & 0.05 &          & Spitzer/MIPS 24 $\mu$m       & 23.513   & 0.000  & 0.05 \\
         & HST/ACS F435W                & 0.4296   & 0.029  & 0.02 &          & Spitzer/MIPS 70 $\mu$m       & 70.389   & 0.000  & 0.10 \\
         & HST/ACS F606W                & 0.5804   & 0.020  & 0.02 &          & Herschel/PACS 100 $\mu$m     & 100.05   & 0.000  & 0.05 \\
         & HST/ACS F775W                & 0.7656   & 0.013  & 0.02 &          & Herschel/PACS 160 $\mu$m     & 159.31   & 0.000  & 0.05 \\
         & HST/ACS F814W                & 0.7979   & 0.012  & 0.02 &          & Herschel/SPIRE 250 $\mu$m    & 247.21   & 0.000  & 0.15 \\
\cline{6-10}
         & HST/ACS F850LP               & 0.8990   & 0.010  & 0.02 & COSMOS   & CFHT/MegaCam $u^{*}$             & 0.3799   & 0.074  & 0.05 \\
         & HST/WFC3 F098M               & 0.9826   & 0.008  & 0.02 &          & Subaru/Suprime-Cam $B$           & 0.4323   & 0.066  & 0.05 \\
         & HST/WFC3 F105W               & 1.0449   & 0.008  & 0.02 &          & Subaru/Suprime-Cam $g^\prime$    & 0.4634   & 0.062  & 0.05 \\
         & HST/WFC3 F125W               & 1.2396   & 0.006  & 0.02 &          & CFHT/MegaCam $g^{\prime}$        & 0.4806   & 0.059  & 0.05 \\
         & HST/WFC3 F160W               & 1.5302   & 0.004  & 0.02 &          & Subaru/Suprime-Cam $V$           & 0.5416   & 0.051  & 0.05 \\
         & VLT/HAWK-I $K_s$             & 2.1403   & 0.002  & 0.05 &          & HST/ACS F606W                    & 0.5804   & 0.046  & 0.02 \\
         & VLT/ISAAC $K_s$              & 2.1541   & 0.002  & 0.05 &          & CFHT/MegaCam $r^{\prime}$        & 0.6189   & 0.041  & 0.05 \\
         & Spitzer/IRAC1                & 3.5314   & 0.001  & 0.05 &          & Subaru/Suprime-Cam $r^\prime$    & 0.6197   & 0.041  & 0.05 \\
         & Spitzer/IRAC2                & 4.4690   & 0.000  & 0.05 &          & CFHT/MegaCam $i^{\prime}$        & 0.7571   & 0.030  & 0.05 \\
         & Spitzer/IRAC3                & 5.6820   & 0.000  & 0.05 &          & Subaru/Suprime-Cam $i^\prime$    & 0.7622   & 0.030  & 0.05 \\
         & Spitzer/IRAC4                & 7.7546   & 0.000  & 0.05 &          & HST/ACS F814W                    & 0.7979   & 0.028  & 0.02 \\
         & Spitzer/MIPS 24 $\mu$m       & 23.513   & 0.000  & 0.05 &          & CFHT/MegaCam $z^{\prime}$        & 0.8782   & 0.024  & 0.05 \\
         & Spitzer/MIPS 70 $\mu$m       & 70.389   & 0.000  & 0.10 &          & Subaru/Suprime-Cam $z^\prime$    & 0.9154   & 0.023  & 0.05 \\
         & Herschel/PACS 100 $\mu$m     & 100.05   & 0.000  & 0.05 &          & VISTA/VIRCAM $Y$                 & 1.0194   & 0.018  & 0.05 \\
         & Herschel/PACS 160 $\mu$m     & 159.31   & 0.000  & 0.05 &          & Mayall/NEWFIRM $J_1$             & 1.0432   & 0.018  & 0.10 \\
         & Herschel/SPIRE 250 $\mu$m    & 247.21   & 0.000  & 0.15 &          & Mayall/NEWFIRM $J_2$             & 1.1922   & 0.014  & 0.10 \\
\cline{1-5}
UDS      & CFHT/MegaCam $u^{*}$             & 0.3799   & 0.091  & 0.05 &          & HST/WFC3 F125W                   & 1.2396   & 0.013  & 0.02 \\
         & Subaru/Suprime-Cam $B$           & 0.4323   & 0.081  & 0.05 &          & VISTA/VIRCAM $J$                 & 1.2497   & 0.013  & 0.05 \\
         & Subaru/Suprime-Cam $V$           & 0.5416   & 0.063  & 0.05 &          & Mayall/NEWFIRM $J_3$             & 1.2757   & 0.013  & 0.10 \\
         & HST/ACS F606W                    & 0.5804   & 0.056  & 0.02 &          & HST/WFC3 F160W                   & 1.5302   & 0.009  & 0.02 \\
         & Subaru/Suprime-Cam $R_c$         & 0.6471   & 0.048  & 0.05 &          & Mayall/NEWFIRM $H_1$             & 1.5578   & 0.009  & 0.10 \\
         & Subaru/Suprime-Cam $i^\prime$    & 0.7622   & 0.037  & 0.05 &          & VISTA/VIRCAM $H$                 & 1.6374   & 0.008  & 0.05 \\
         & HST/ACS F814W                    & 0.7979   & 0.034  & 0.02 &          & Mayall/NEWFIRM $H_2$             & 1.7041   & 0.008  & 0.10 \\
         & Subaru/Suprime-Cam $z^\prime$    & 0.9154   & 0.028  & 0.05 &          & VISTA/VIRCAM $K_s$               & 2.1408   & 0.006  & 0.05 \\
         & VLT/HAWK-I $Y$                   & 1.0187   & 0.023  & 0.05 &          & Mayall/NEWFIRM $K$               & 2.1639   & 0.006  & 0.10 \\
         & HST/WFC3 F125W                   & 1.2396   & 0.016  & 0.02 &          & Spitzer/IRAC1                    & 3.5314   & 0.003  & 0.05 \\
         & UKIRT/WFCAM $J$                  & 1.2521   & 0.016  & 0.05 &          & Spitzer/IRAC2                    & 4.4690   & 0.000  & 0.05 \\
         & HST/WFC3 F160W                   & 1.5302   & 0.011  & 0.02 &          & Spitzer/IRAC3                    & 5.6820   & 0.000  & 0.05 \\
         & UKIRT/WFCAM $H$                  & 1.6406   & 0.010  & 0.05 &          & Spitzer/IRAC4                    & 7.7546   & 0.000  & 0.05 \\
         & VLT/HAWK-I $K_s$                 & 2.1403   & 0.007  & 0.05 &          & Spitzer/MIPS 24 $\mu$m           & 23.513   & 0.000  & 0.05 \\
         & UKIRT/WFCAM $K$                  & 2.2261   & 0.007  & 0.05 &          & Spitzer/MIPS 70 $\mu$m           & 70.389   & 0.000  & 0.10 \\
         & Spitzer/IRAC1                    & 3.5314   & 0.004  & 0.05 &          & Herschel/PACS 100 $\mu$m         & 100.05   & 0.000  & 0.05 \\
         & Spitzer/IRAC2                    & 4.4690   & 0.000  & 0.05 &          & Herschel/PACS 160 $\mu$m         & 159.31   & 0.000  & 0.05 \\
         & Spitzer/IRAC3                    & 5.6820   & 0.000  & 0.05 &          & Herschel/SPIRE 250 $\mu$m        & 247.21   & 0.000  & 0.15 \\
         & Spitzer/IRAC4                    & 7.7546   & 0.000  & 0.05 &          &    &    &   &  \\
         & Spitzer/MIPS 24 $\mu$m           & 23.513   & 0.000  & 0.05 &          &    &    &   &  \\
         & Spitzer/MIPS 70 $\mu$m           & 70.389   & 0.000  & 0.10 &          &    &    &   &  \\
         & Herschel/PACS 100 $\mu$m         & 100.05   & 0.000  & 0.05 &          &    &    &   &  \\
         & Herschel/PACS 160 $\mu$m         & 159.31   & 0.000  & 0.05 &          &    &    &   &  \\
         & Herschel/SPIRE 250 $\mu$m        & 247.21   & 0.000  & 0.15 &          &    &    &   &  \\
\enddata
\tablenotetext{a}{Mean wavelength of the filter calculated as $\lambda_{\textrm{mean}} = \frac{\int \lambda T(\lambda) d\lambda}{\int T(\lambda) d\lambda}$, where $T(\lambda)$ is the filter transmission function.}
\tablenotetext{b}{Galactic extinction for the center of the field.}
\tablenotetext{c}{Calibration uncertainties as given by the corresponding instrument user handbook.}
\end{deluxetable*}

\subsection{SINGS/KINGFISH sample} \label{sec:KINGFISH}

We supplemented our CANDELS sample with an additional 18 local disk-dominated galaxies from the combined SINGS and KINGFISH sample given in \citet{2017ApJ...837...90D}, since UV star formation tracers are also commonly used in local galaxies. We first selected galaxies to be star-forming spiral galaxies (Sa and later types) as given by their optical morphologies in \citet{2017ApJ...837...90D}. They were also selected to not be AGN-dominated (i.e., Seyfert galaxies) to limit any contamination of the photometry by AGNs, using the nuclear type given in \citet{2003PASP..115..928K}. Further, we excluded galaxies with low Galactic latitude (absolute latitude $<15^\circ$), as the large number of foreground stars can result in nonnegligible contamination of the observed fluxes. We also excluded any galaxies that are known to be or have companion galaxies (e.g., NGC 1097 and NGC 5457), as the interaction between companions could impact disk morphology, resulting in distorted inclination estimates. Finally, we visually inspected images of the remaining galaxies and excluded any that are irregularly shaped or contain bright or dominant bulges. With these criteria, our SINGS/KINGFISH sample includes the following 18 galaxies: NGC 24, NGC 337, NGC 628, NGC 925, NGC 2403, NGC 2976, NGC 3049, NGC 3184, NGC 3198, NGC 3938, NGC 4236, NGC 4254, NGC 4536, NGC 4559, NGC 4631, NGC 5055, NGC 7331, and NGC 7793.

The photometry that we used for the SINGS/KINGFISH sample was derived by \citet{2017ApJ...837...90D} and is given in their Table 2. We corrected this photometry for Galactic extinction using the $E(B-V)$ values quoted in \citet{2017ApJ...837...90D} along with their $A_V$-normalized extinction values by bandpass. These extinction values were derived from the \citet{2011ApJ...737..103S} recalibration of the \citet{1998ApJ...500..525S} dust maps and assuming a \citet{2001ApJ...554..778L} reddening curve with $R_V = 3.1$. Unlike the CANDELS sample, we do not add any additional fractional calibration uncertainties to these flux uncertainties, as fractional calibration uncertainties are already included in the uncertainties given by \citet{2017ApJ...837...90D}.

The axis ratios for the SINGS/KINGFISH sample were gathered for each galaxy from the HyperLeda database\footnote{\url{http://leda.univ-lyon1.fr/}} \citep{2014A&A...570A..13M}. We do not use the major and minor axis values quoted in \citet{2017ApJ...837...90D} for our axis ratios, as they were chosen to encapsulate practically all of the fluxes at all measured wavelengths. Instead, the HyperLeda axis ratios and their uncertainties are derived from 25 mag/arcsec$^2$ $B$-band isophotes, which is more consistent with the axis ratio derivation of the CANDELS sample.

\section{Derivation of Physical Properties} \label{sec:Derive}

\subsection{Lightning SED Modeling} \label{sec:Lightning}

We fitted the corrected photometry (as discussed in Section~\ref{sec:Data}) of each galaxy using the SED fitting code \texttt{Lightning}\footnote{Version 2.0: \url{https://github.com/rafaeleufrasio/lightning}} \citep{2017ApJ...851...10E,2021ApJ...923...26D}, assuming a 10\% model uncertainty for each band. For the fits, we assumed the same model as \citet{2021ApJ...923...26D} when fitting using the inclination-dependent model with an image-based inclination prior. This model consists of an SFH that has five constant SFR age bins, the inclination-dependent attenuation curves described in \citet{2021ApJ...923...26D}, and the dust emission of \citet{2007ApJ...657..810D}. A full description of the model, a list of all free parameters and their corresponding prior distributions, and a description of the inclination-dependent attenuation curves can be found in Section~5, Table~2, and Section~4.3 of \citet{2021ApJ...923...26D}, respectively. The only change to the model occurred for the SINGS/KINGFISH sample, where the lower limit of $U_{\rm min}$ (the minimum value of the radiation field intensity $U$ for the dust emission) was changed from 0.7 to 0.1, since the SINGS/KINGFISH sample has rest-frame submillimeter data. For the image-based inclination prior distributions, we derived probability distributions of inclination given our axis ratios via the Monte Carlo method presented in Section~3 of \citet{2021ApJ...923...26D}. The method creates a distribution of inclination for a given galaxy that accounts for variation in the measured axis ratio due to galaxy intrinsic thickness and asymmetry.

Using this model, we fitted the SED of each galaxy using the adaptive Markov Chain Monte Carlo (MCMC) procedure in \texttt{Lightning}. We ran each MCMC fit for $2\times 10^5$ iterations and tested for convergence of the chains to a best solution using 10 parallel chains, each started at random starting locations within the parameter ranges. Convergence was tested using the Gelman--Rubin test \citep{Gelman1992,Brooks1998} on the last 5000 iterations of the parallel chains, which indicated that the set of parallel chains for all galaxies converged to the same solution (i.e., $\sqrt{\hat{R}} \approx 1$). For each galaxy, we then used the last 5000 iterations of the parallel chain with the minimum $\chi^2$ for our output parameter distributions. Finally, using the minimum $\chi^2$ of each galaxy, we tested how well our model described the data by performing a $\chi^2$ goodness-of-fit test. The results of this test showed a relatively flat $P_{\rm null}$ distribution, which indicates that the model has acceptably fit the SEDs.

\subsection{Derived Physical Properties} \label{sec:PhysProp}

\begin{deluxetable*}{lccccccccc}
\tabletypesize{\footnotesize}
\tablecaption{Galaxy Sample and Properties. \label{table:sampledata}}
\tablecolumns{10}
\tablehead{ Name & R.A.  & Decl. & $D$ & $z$ & $q$ & $1-\cos i$ & $L_{\rm FUV}$ & $A_{\rm FUV}$ & \\ & (deg) & (deg) & (Mpc) & & & & ($L_\odot \ {\rm Hz}^{-1}$) & (mag) & \\(1) & (2) & (3) & (4) & (5) & (6) & (7) & (8) & (9) & }
\startdata
J123624.82+620719.2 & 189.1034 & $ 62.1220$ & $\cdots$ & 0.11 & $0.591 \pm 0.001$ & $0.465 \pm 0.096$ & $(1.471 \pm 0.381) \times 10^{-6}$ & $0.753 \pm 0.234$ & $\cdots$ \\
J123723.47+621448.3 & 189.3478 & $ 62.2468$ & $\cdots$ & 0.25 & $0.595 \pm 0.034$ & $0.513 \pm 0.094$ & $(1.252 \pm 0.294) \times 10^{-6}$ & $1.346 \pm 0.270$ & $\cdots$ \\
J123654.64+621127.1 & 189.2277 & $ 62.1909$ & $\cdots$ & 0.25 & $0.118 \pm 0.001$ & $0.932 \pm 0.026$ & $(2.106 \pm 0.585) \times 10^{-7}$ & $3.397 \pm 0.280$ & $\cdots$ \\
J123733.50+621941.0 & 189.3896 & $ 62.3281$ & $\cdots$ & 0.27 & $0.243 \pm 0.236$ & $0.824 \pm 0.041$ & $(2.591 \pm 0.800) \times 10^{-7}$ & $3.653 \pm 0.326$ & $\cdots$ \\
J123809.19+621638.1 & 189.5383 & $ 62.2772$ & $\cdots$ & 0.28 & $0.136 \pm 0.052$ & $0.925 \pm 0.031$ & $(5.536 \pm 1.342) \times 10^{-7}$ & $3.124 \pm 0.230$ & $\cdots$ \\
J123711.77+621514.9 & 189.2990 & $ 62.2541$ & $\cdots$ & 0.30 & $0.583 \pm 0.037$ & $0.464 \pm 0.095$ & $(3.351 \pm 0.651) \times 10^{-6}$ & $0.835 \pm 0.207$ & $\cdots$ \\
J123745.89+621435.0 & 189.4412 & $ 62.2430$ & $\cdots$ & 0.30 & $0.211 \pm 0.057$ & $0.894 \pm 0.027$ & $(1.204 \pm 0.241) \times 10^{-7}$ & $4.835 \pm 0.209$ & $\cdots$ \\
J123615.96+621008.2 & 189.0665 & $ 62.1689$ & $\cdots$ & 0.34 & $0.500 \pm 0.002$ & $0.633 \pm 0.081$ & $(1.883 \pm 0.446) \times 10^{-6}$ & $2.404 \pm 0.300$ & $\cdots$ \\
J123654.12+621737.8 & 189.2255 & $ 62.2938$ & $\cdots$ & 0.38 & $0.537 \pm 0.024$ & $0.670 \pm 0.068$ & $(1.691 \pm 0.383) \times 10^{-6}$ & $2.575 \pm 0.286$ & $\cdots$ \\
J123701.67+621814.4 & 189.2570 & $ 62.3040$ & $\cdots$ & 0.44 & $0.472 \pm 0.069$ & $0.752 \pm 0.049$ & $(1.025 \pm 0.234) \times 10^{-6}$ & $3.101 \pm 0.289$ & $\cdots$ \\
J123726.54+621826.3 & 189.3606 & $ 62.3073$ & $\cdots$ & 0.44 & $0.555 \pm 0.048$ & $0.531 \pm 0.093$ & $(1.803 \pm 0.293) \times 10^{-6}$ & $1.473 \pm 0.231$ & $\cdots$ \\
J123630.86+621433.5 & 189.1286 & $ 62.2426$ & $\cdots$ & 0.44 & $0.497 \pm 0.045$ & $0.595 \pm 0.086$ & $(1.239 \pm 0.214) \times 10^{-6}$ & $1.867 \pm 0.245$ & $\cdots$ \\
J123743.50+621631.7 & 189.4312 & $ 62.2755$ & $\cdots$ & 0.44 & $0.212 \pm 0.137$ & $0.886 \pm 0.046$ & $(1.635 \pm 0.238) \times 10^{-6}$ & $2.376 \pm 0.172$ & $\cdots$ \\
J123654.16+620821.4 & 189.2257 & $ 62.1393$ & $\cdots$ & 0.45 & $0.282 \pm 0.005$ & $0.804 \pm 0.048$ & $(2.716 \pm 0.731) \times 10^{-7}$ & $3.424 \pm 0.343$ & $\cdots$ \\
J123653.60+622111.6 & 189.2233 & $ 62.3532$ & $\cdots$ & 0.47 & $0.231 \pm 0.095$ & $0.892 \pm 0.063$ & $(2.570 \pm 0.378) \times 10^{-6}$ & $2.106 \pm 0.172$ & $\cdots$ \\
$\cdots$ & $\cdots$ & $\cdots$ & $\cdots$ & $\cdots$ & $\cdots$ & $\cdots$ & $\cdots$ & $\cdots$ & $\cdots$ \\
NGC 0024 &   2.4829 & $-24.9653$ &  8.20 & $\cdots$ & $0.389 \pm 0.025$ & $0.670 \pm 0.090$ & $(1.686 \pm 0.207) \times 10^{-7}$ & $0.785 \pm 0.139$ & $\cdots$ \\
NGC 0337 &  14.9613 & $ -7.5789$ & 19.30 & $\cdots$ & $0.647 \pm 0.057$ & $0.378 \pm 0.096$ & $(1.088 \pm 0.084) \times 10^{-6}$ & $1.732 \pm 0.100$ & $\cdots$ \\
NGC 0628 &  24.1767 & $ 15.7864$ &  7.20 & $\cdots$ & $0.944 \pm 0.085$ & $0.123 \pm 0.080$ & $(1.170 \pm 0.117) \times 10^{-6}$ & $0.987 \pm 0.166$ & $\cdots$ \\
NGC 0925 &  36.8067 & $ 33.5844$ &  9.12 & $\cdots$ & $0.537 \pm 0.038$ & $0.498 \pm 0.085$ & $(1.196 \pm 0.113) \times 10^{-6}$ & $0.776 \pm 0.113$ & $\cdots$ \\
NGC 2403 & 114.2296 & $ 65.5928$ &  3.50 & $\cdots$ & $0.505 \pm 0.042$ & $0.542 \pm 0.079$ & $(8.678 \pm 0.928) \times 10^{-7}$ & $0.985 \pm 0.100$ & $\cdots$ \\
$\cdots$ & $\cdots$ & $\cdots$ & $\cdots$ & $\cdots$ & $\cdots$ & $\cdots$ & $\cdots$ & $\cdots$ & $\cdots$ \\
\enddata
\tablecomments{The full version of this table contains 20 columns of information for all galaxies in our CANDELS and SINGS/KINGFISH samples. An abbreviated version of the table is displayed here to illustrate its form and content. Col.(1): Adopted galaxy designation. Col.(2): Right ascension  in J2000. Col.(3): Declination in J2000. Col.(4): Adopted distance (only for SINGS/KINGFISH sample). Col.(5): Adopted spectroscopic redshift (only for CANDELS sample). Col.(6): Measured axis ratio. Col.(7): Inclination derived from \texttt{Lightning}. Col.(8): Attenuated model rest-frame FUV-band luminosity in terms of $L_\nu$. Col.(9): FUV-band attenuation. Col.(10): Attenuated model rest-frame NUV-band luminosity in terms of $L_\nu$. Col.(11): NUV-band attenuation. Col.(12): Attenuated model rest-frame WFC3/F275W-band luminosity in terms of $L_\nu$. Col.(13): WFC3/F275W-band attenuation. Col.(14-17): Attenuated model rest-frame J, H, K, and 3.6-band luminosities in terms of $L_\nu$, respectively. Col.(18): Total integrated infrared luminosity. Col.(19): Total stellar mass. Col.(20): Recent star formation rate of last 100 Myr. \\(This table is available in its entirety in machine-readable form.)}
\end{deluxetable*}

From the output parameter distributions of the SED fitting, we derived the various properties needed for our analysis (e.g., inclination, $L_{\rm FUV}$, $A_{\rm FUV}$, $L_{\rm TIR}$, etc.). All of these properties for our sample are given in Table~\ref{table:sampledata}. For the bandpass luminosities (calculated as $L_\nu$), they were derived by convolving the corresponding filter transmission function with the attenuated rest-frame model spectrum to avoid any redshift dependencies. Additionally, isotropy was assumed when calculating these luminosities from the model spectra, since isotropy is typically assumed when converting observed fluxes to luminosities. We note that for the remainder of the paper, when we refer to any attenuated (or unattenuated) bandpass luminosity or color, we are implicitly referring to these rest-frame model luminosities as given in Table~\ref{table:sampledata}. From the properties given in Table~\ref{table:sampledata}, we derived four additional properties needed for our analysis, specifically, $a_{\rm corr}$, $\beta$, $\beta_0$, and $a_\beta$ (see Equations~\ref{eq:acorr} and~\ref{eq:AUVbeta}). A detailed description of how we calculated these properties is given below.

\begin{figure}[t!]
\centerline{
\includegraphics[width=8.75cm]{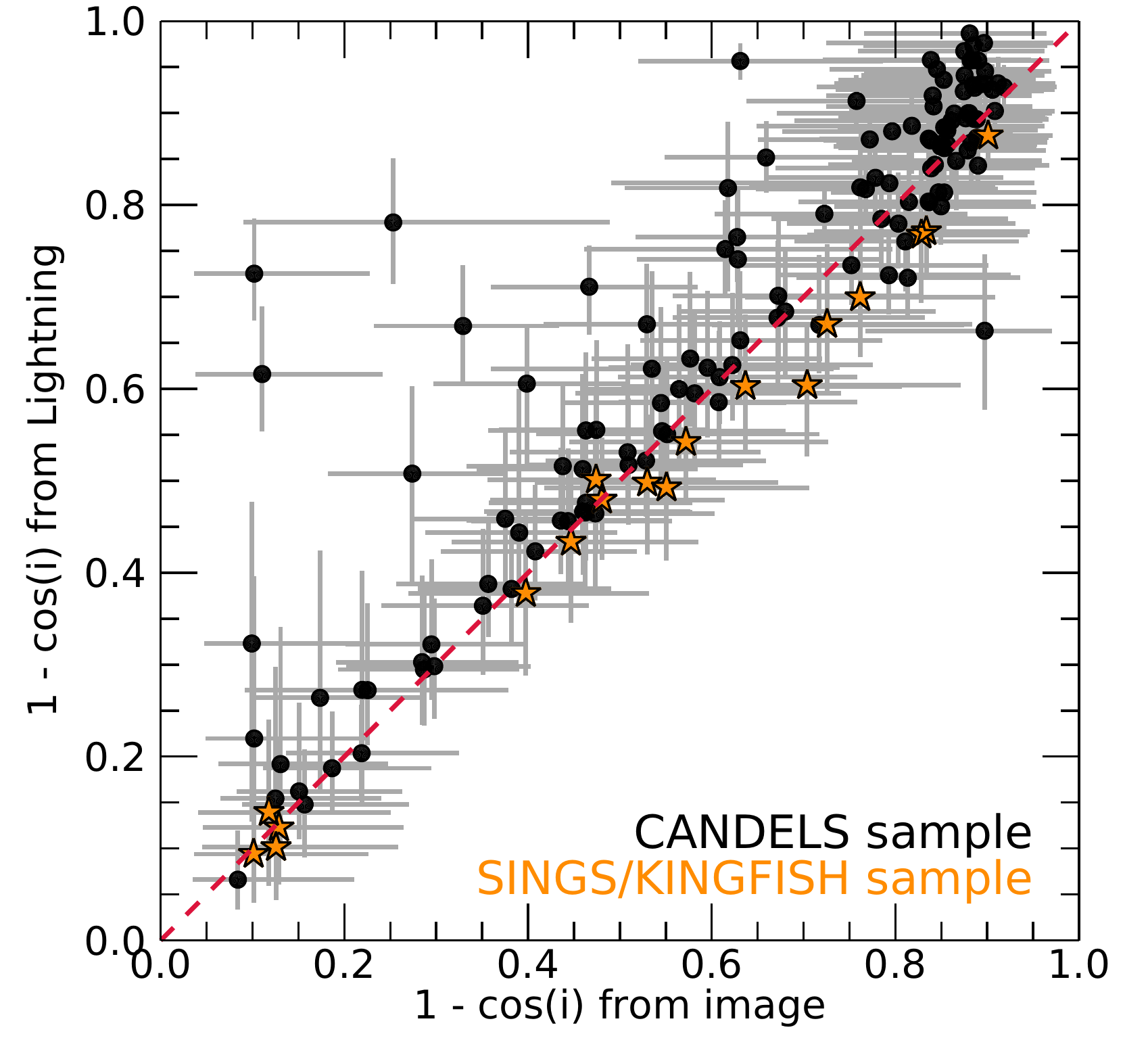}
}
\caption{
Inclinations derived from \texttt{Lightning} vs. the image-based inclinations derived from the axis ratio using the Monte Carlo method of \citet{2021ApJ...923...26D}. The black circles are the inclination estimates for the CANDELS sample of galaxies, and the orange stars are the inclination estimates for the local SINGS/KINGFISH sample of galaxies. All of the SINGS/KINGFISH inclinations and the vast majority of CANDELS inclinations fall along the one-to-one line, indicating that the image-based inclination priors are informative.
}
\label{fig:incvinc}
\end{figure}

To first asses the accuracy of our derived inclinations, we compared these inclinations to the image-based inclination priors derived from the axis ratios. We show this comparison in Figure~\ref{fig:incvinc}, where the vast majority of galaxies fall along the one-to-one line. However, the small number of galaxies that deviate significantly from the one-to-one line are all from the CANDELS sample. \citet{2021ApJ...923...26D} discussed that the galaxies far from the one-to-one line may have disks that are significantly thicker and dynamically hotter than the galaxies in the local universe, on which the inclination-dependent model was based. Therefore, the inclination-dependent model may not be physically appropriate for these galaxies. However, we continued to use our inclinations derived from \texttt{Lightning} as our inclination estimates and did not remove those four to five galaxies from our sample, as they had a statistically insignificant impact on our results.

\begin{figure}[t!]
\centerline{
\includegraphics[width=8.75cm]{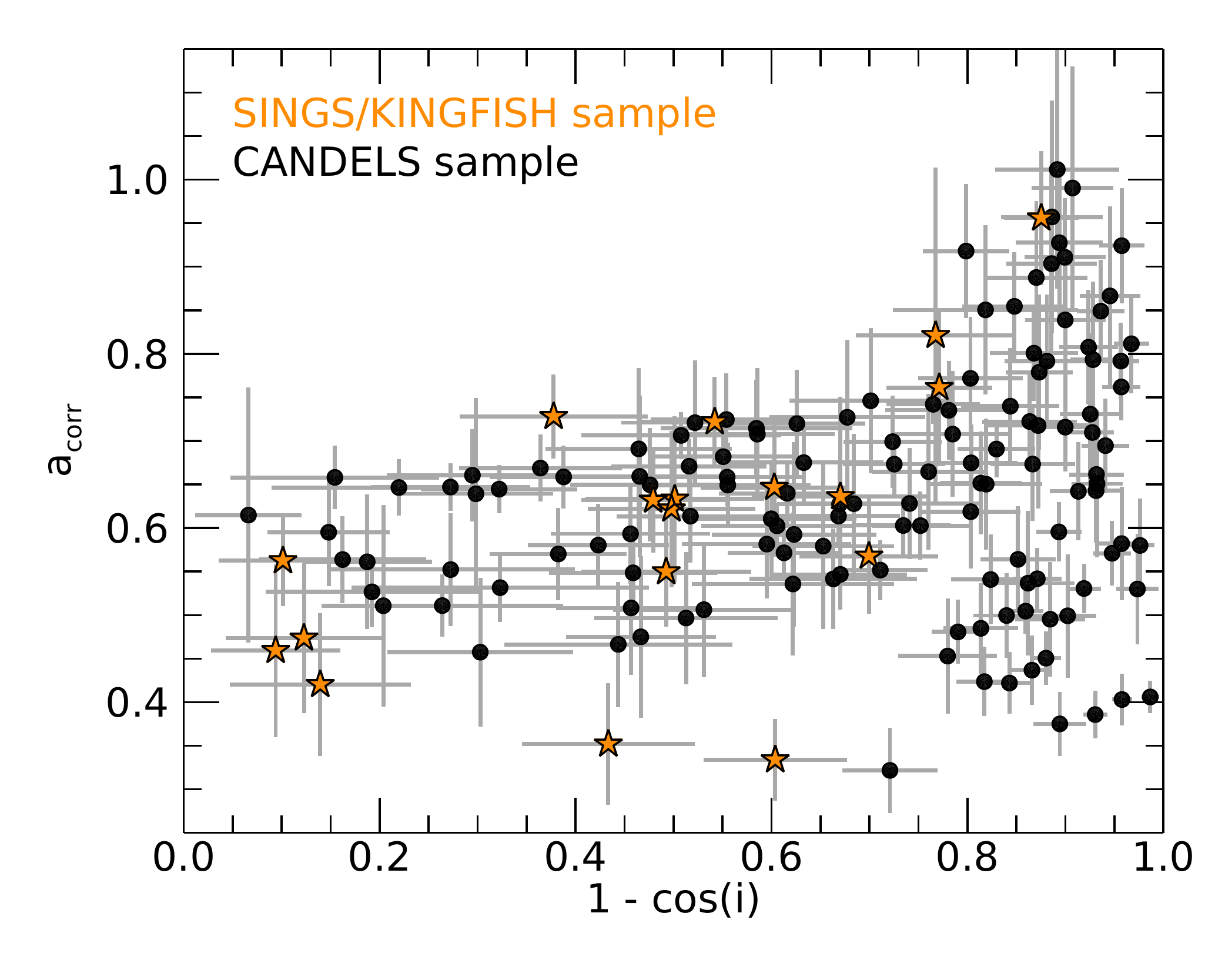}
}
\caption{
Inclinations derived from \texttt{Lightning} vs. $a_{\rm corr}$. The black circles represent the CANDELS sample of galaxies, and the orange stars represent the local SINGS/KINGFISH sample of galaxies. As inclination increases from face-on to edge-on, the value of $a_{\rm corr}$ tends to increase as expected. However, edge-on galaxies have a wider variation compared to face-on galaxies due to the variation in each galaxy's physical properties.
}
\label{fig:incvacorr}
\end{figure}

To derive $a_{\rm corr}$ (see Equation~\ref{eq:acorr}), we utilized the attenuated and unattenuated rest-frame model FUV luminosities along with the model $L_{\rm TIR}$. After converting the FUV luminosities to monochromatic luminosities (i.e., $\nu L_\nu$), $a_{\rm corr}$ was calculated following Equation~\ref{eq:acorr}. Figure~\ref{fig:incvacorr} shows how $a_{\rm corr}$ varies with inclination. Typically, as inclination increases from face-on to edge-on, the value of $a_{\rm corr}$ increases as expected. However, edge-on galaxies have a broad range of $a_{\rm corr}$ values, with some having lower $a_{\rm corr}$ values compared to face-on galaxies. As will be discussed in Section~\ref{sec:acorr}, this variation at high inclinations is correlated to the variation in each galaxy's physical properties, specifically the specific SFR (sSFR; defined as the SFR divided by stellar mass).

\begin{figure*}[t!]
\centerline{
\includegraphics[width=18cm]{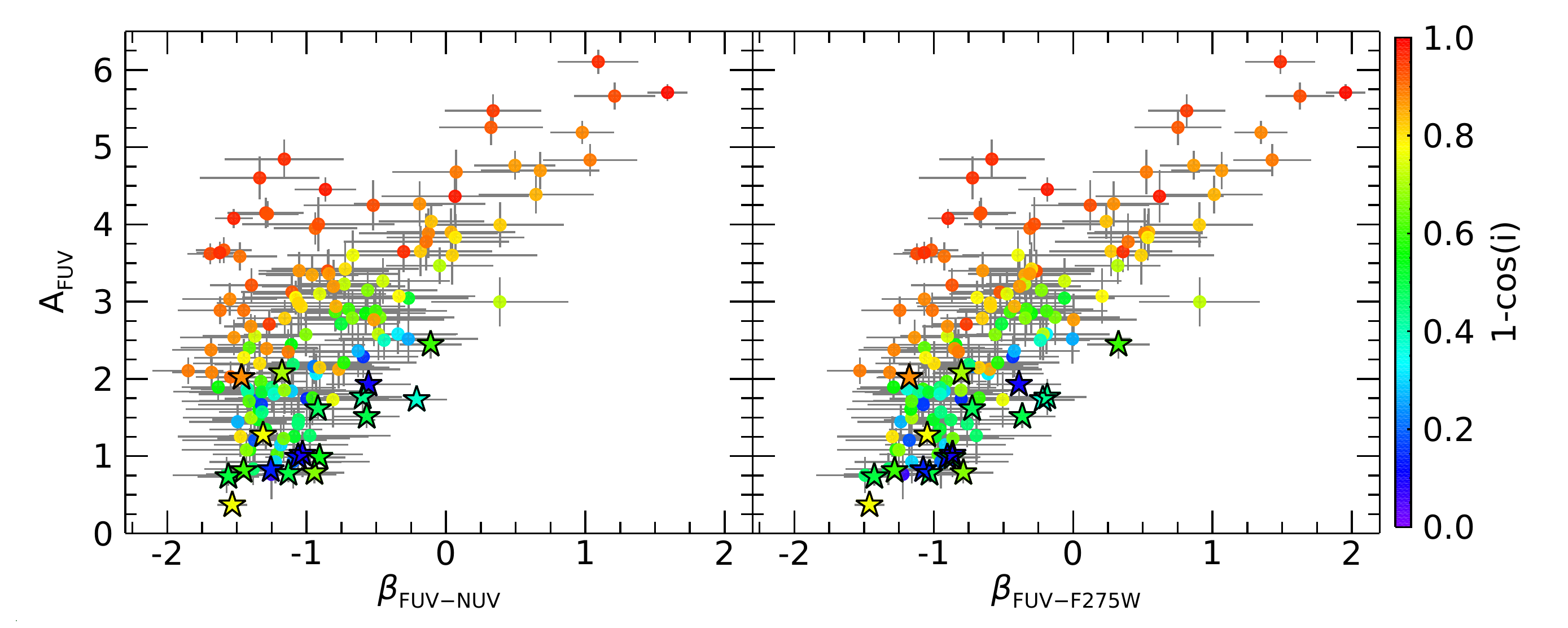}
}
\caption{
Shown is $A_{\rm FUV}$ vs. $\beta$ for the galaxies in our sample, with the right panel $\beta$ being calculated using the rest-frame model FUV and NUV bands ($\beta_{\rm FUV-NUV}$), and the left panel $\beta$ being calculated using the rest-frame model FUV and F275W bands ($\beta_{\rm FUV-F275W}$). The circles are the galaxies in the CANDELS sample, and the stars are the galaxies in the SINGS/KINGFISH sample. Both are colored based on their inclination as derived by \texttt{Lightning}. A clear transition can be seen in $A_{\rm FUV}$ as inclination increases for a fixed value of $\beta$.
}
\label{fig:afuvbeta}
\end{figure*}

Following the procedures of past studies, where observations in only two UV bands are typically available, we derive the UV slope $\beta$ from
\begin{equation} \label{eq:beta}
\beta=\frac{\log_{10}(L_{\nu, 1}/L_{\nu, 2})}{\log_{10}(\lambda_1/\lambda_2)}-2,
\end{equation}
where $L_\nu$ is the attenuated rest-frame model luminosities for two UV bandpasses\footnote{For observations, the fluxes ($F_\nu$) can simply be swapped for the luminosities, since isotropic luminosities have the property of $L_\nu \propto F_\nu$.}, and $\lambda$ is the corresponding central wavelength of the bandpasses. To calculate $\beta_0$, the attenuated rest-frame model luminosities in Equation~\ref{eq:beta} can simply be swapped for the unattenuated rest-frame model luminosities, since $\beta_0$ is an intrinsic, dust-free property.

To derive $a_\beta$, we substituted Equation~\ref{eq:beta} for both $\beta$ and $\beta_0$ into Equation~\ref{eq:AUVbeta} along with $A_\lambda=-2.5 \log_{10}(L_\nu/L_{\nu, 0})$. For the FUV-band attenuation ($A_{\rm FUV}$), this gives
\begin{equation}
a_\beta = \frac{A_{\rm FUV}\log_{10}(\lambda_1/\lambda_2)}{0.4(A_{\lambda, 2}-A_{\lambda, 1})},
\label{eq:abeta}
\end{equation}
where $A_{\lambda, i}$ is the attenuation for the $i$th UV bandpass at $\lambda_i$ in Equation~\ref{eq:beta}. From Equation~\ref{eq:abeta}, $a_\beta$ can be seen to depend primarily on the attenuation curve, but it additionally depends on the choice of UV bandpasses. This same UV bandpass dependence is also present in Equation~\ref{eq:beta} for $\beta$ (and similarly $\beta_0$), and it can have a significant impact on the derived values of both $\beta$ and $a_\beta$. For example, if one of the selected UV bandpasses contains the rest-frame 2175~\AA\  bump feature, which is present in our attenuation curves, then the measurements of $\beta$ will be biased to smaller, more negative values \citep{2005MNRAS.360.1413B,2009ApJ...706..553B,2010ApJ...718..184C,2011MNRAS.417.1760W,2013ApJ...775L..16K,2017ApJ...851...90B,2017MNRAS.472.2315P,2018MNRAS.475.2363T} and $a_\beta$ to larger values. 

Since rest-frame observations that avoid the UV bump are not always available, we calculated two sets of values for $\beta$, $\beta_0$, and $a_\beta$ via Equations~\ref{eq:beta} and~\ref{eq:abeta}. The first set includes the rest-frame model GALEX FUV ($\lambda = 1530$~\AA) and near-UV (NUV; $\lambda = 2260$~\AA) bandpasses, with the NUV bandpass overlapping with the UV bump. This set and subsequent relations derived in Section~\ref{sec:M99} will be more applicable to galaxies that have observational bands that contain the rest-frame UV bump feature ($\sim2175$ \AA). As for the second set, we used the rest-frame model GALEX FUV and HST WFC3/F275W ($\lambda = 2690$~\AA) bandpasses, both of which avoid the bump feature. The choice of the WFC3/F275W band is motivated by \citet{2017MNRAS.472.2315P}, who showed that the WFC3/F275W band has minimal overlap with the UV bump, and, when used in combination with the GALEX FUV, calculated values of $\beta$ are minimally impacted by the UV bump feature. Therefore, this set will be applicable to galaxies whose observations are relatively free of any bump feature contamination. 

Figure~\ref{fig:afuvbeta} shows $A_{\rm FUV}$ (derived from the SED fits) versus both sets of $\beta$ for the galaxies in our sample, with each galaxy being colored by its inclination derived from \texttt{Lightning}. The values of $\beta$ in the left panel, which were derived from Equation~\ref{eq:beta} using the FUV and NUV bands, can be seen to be more negative than those in the right panel, which were derived with the FUV and F275W bands. Additionally, a clear inclination dependence can be seen in $A_{\rm FUV}$ for a fixed value of $\beta$. This variation with inclination is caused by $a_\beta$, the shape of the attenuation curve, being inclination-dependent.

Figure~\ref{fig:incvabeta} shows how the two sets of $a_{\beta}$ vary with inclination. The orange circles and stars represent the CANDELS and SINGS/KINGFISH sample of galaxies, respectively, whose $a_\beta$ values were derived using the FUV and NUV bands. The blue circles and stars represent the CANDELS and SINGS/KINGFISH sample of galaxies, respectively, whose $a_\beta$ values were derived using the FUV and F275W bands. Both sets show an expected trend of increasing with inclination, but the values of $a_\beta$ derived using the NUV band can clearly be seen to have larger values compared to those using the F275W band. These larger values of $a_\beta$ are due to the UV bump, the presence of which causes an increase in attenuation in the NUV. The scatter that is present in both sets of $a_\beta$ values is due to other attenuation parameters (i.e., the face-on optical depth in the $B$ band, $\tau_B^f$, and the galaxy clumpiness factor, $F$) influencing the value of $a_\beta$. The value of $\tau_B^f$ can also affect the strength of the UV bump, which causes larger scatter by approximately a factor of 2 at all inclinations in the values of $a_\beta$ derived using the NUV band compared to those using the F275W band.

\subsection{Simulated Data} \label{sec:SimData}

\begin{figure}[t!]
\centerline{
\includegraphics[width=8.75cm]{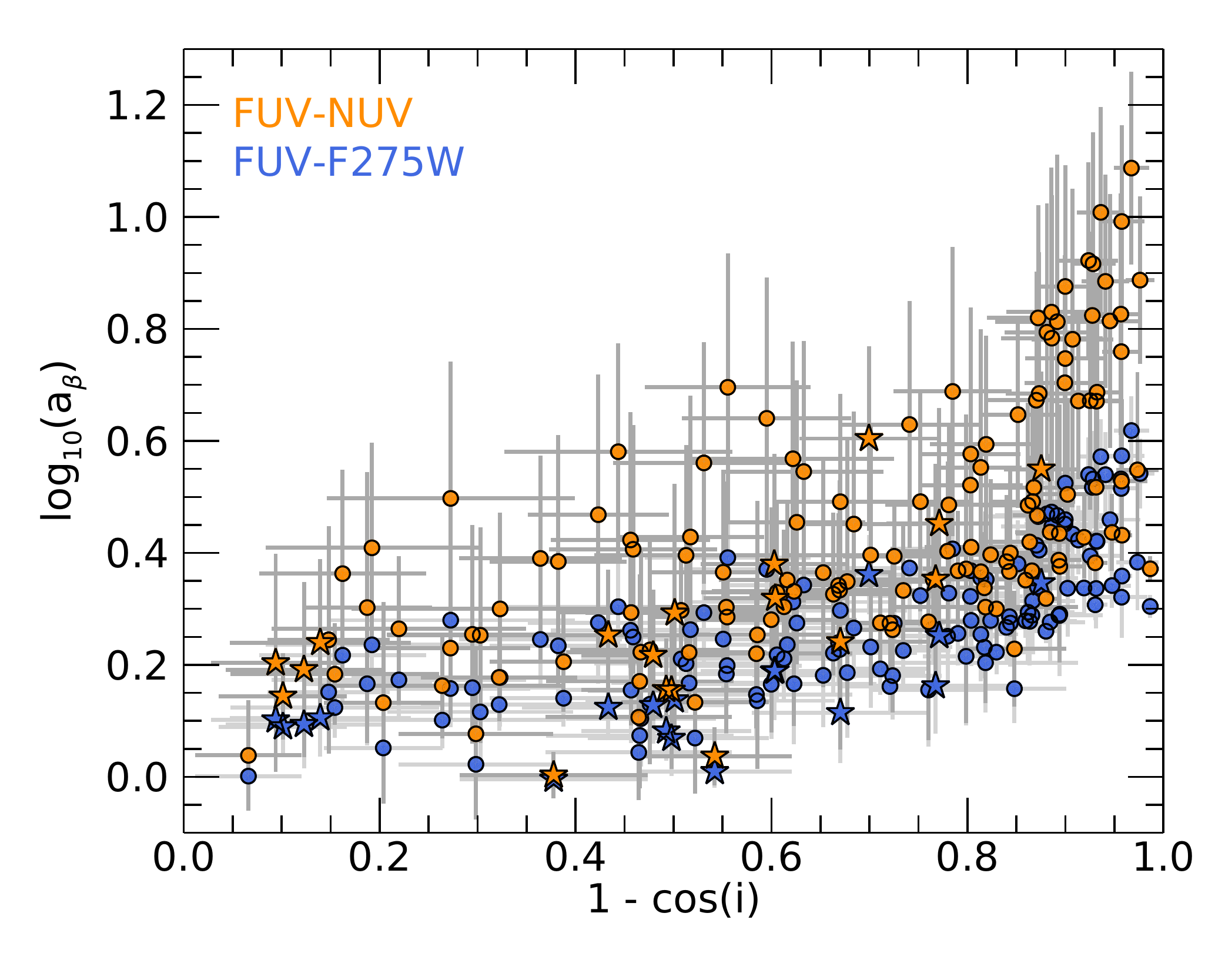}
}
\caption{
Inclinations derived from \texttt{Lightning} vs. $a_\beta$. The orange circles and stars represent the CANDELS and SINGS/KINGFISH samples of galaxies, respectively, whose $a_\beta$ values were derived using the rest-frame model FUV and NUV bands. The blue circles and stars represent the CANDELS and SINGS/KINGFISH samples of galaxies, respectively, whose $a_\beta$ values were derived using the rest-frame model FUV and F275W bands. The difference between sets of $a_\beta$ values is due to the NUV band being contaminated by the 2175~\AA\ bump feature, which biases $a_\beta$ to higher values. The scatter that is present in both sets of $a_\beta$ values is due to other attenuation parameters besides inclination influencing the value of $a_\beta$.
}
\label{fig:incvabeta}
\end{figure}
\begin{figure*}[t!]
\centerline{
\includegraphics[width=18cm]{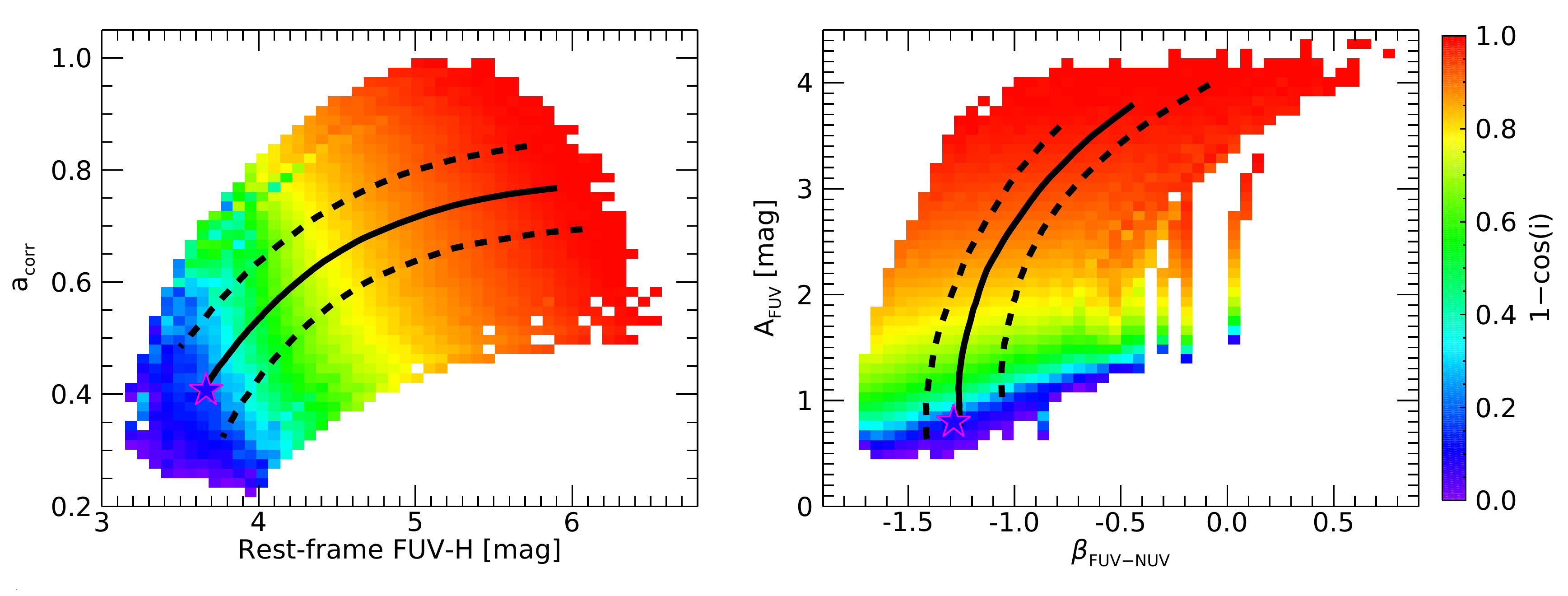}
}
\caption{
(Left) $a_{\rm corr}$ vs. rest-frame model FUV--$H$ color. (Right) $A_{\rm FUV}$ vs. $\beta$ calculated using the rest-frame model FUV and NUV bands ($\beta_{\rm FUV-NUV}$). Each panel shows the simulated data for NGC 3184. The rainbow background image in each panel is the averaged inclination of the simulated data points contained within each pixel. The solid (dashed) black lines show the median (1$\sigma$ spread) of each parameter distribution for each inclination grid point, and the magenta star is the best-fit data point from the original parameter distribution chains. In each panel, the rainbow transition indicates how each parameter changes in parameter space with inclination.
}
\label{fig:simdata}
\end{figure*}

As can be inferred from Figures~\ref{fig:incvred},~\ref{fig:incvinc},~\ref{fig:incvacorr}, and~\ref{fig:incvabeta}, our sample of galaxies does not have an expected randomly selected distribution in inclination (uniform in $1-\cos i$ space), instead having more highly inclined galaxies compared to nearly face-on galaxies. This bias is due to the visual inspection process in our sample selection, since edge-on galaxies are less likely to be confused for irregular galaxies compared to face-on spirals. To more fully sample inclination space and better quantify inclination-dependent trends in $a_{\rm corr}$ and the $A_{\rm FUV}$--$\beta$ relation in Sections~\ref{sec:acorrRel} and~\ref{sec:M99Rel}, respectively, we simulated how all galaxies in our sample would appear if observed over a full range of possible inclinations. To achieve this, we used our solutions for the SFHs of our galaxies, along with our inclination-dependent attenuation curves, to construct emergent rest-frame SEDs of our galaxies across a grid of inclinations. Thus, these simulated models allow for our sample's variety of SFHs to be available at all inclinations, rather than the SFHs being limited to the corresponding measured inclination of each galaxy.

To generate the simulated data for a given galaxy, we utilized the output parameter distributions (i.e., the resulting 5000 element Markov chain of each parameter) of the SED fitting. For a given element in the chain, all parameters excluding inclination were fixed, and attenuated rest-frame models were generated for a grid of inclinations (0--1 in steps of 0.01 in $\cos i$ space). From these attenuated models, the necessary physical properties for our study (e.g., $L_{\rm FUV}$, $A_{\rm FUV}$, $a_{\rm corr}$, $\beta$, etc.) were derived and recorded. This process was performed for all 5000 elements in the chain and, subsequently, each galaxy in the sample. Therefore, the simulated data set for a given physical property consists of a unique distribution for each galaxy in our sample at each inclination grid point. We note that, since inclination only affects attenuation, the unattenuated stellar models did not need to be simulated, as they would be the same at all inclinations.

An example of the simulated data for the randomly selected SINGS/KINGFISH galaxy NGC 3184 is displayed in Figure~\ref{fig:simdata}. For both panels, the background rainbow image is the averaged inclination of the simulated data points contained within each pixel. These images show how the distribution of each parameter changes as inclination is varied from face-on to edge-on, with the solid (dashed) black lines showing the median (1$\sigma$ spread) of each parameter distribution for each inclination grid point. The left panel shows a clear transition to larger values of $a_{\rm corr}$ and rest-frame FUV--$H$ color (the reason for using color is discussed in Section~\ref{sec:acorrProp}) as inclination increases. As for the right panel, which shows $A_{\rm FUV}$ versus $\beta$, $A_{\rm FUV}$ transitions to large values with inclination as expected. While $\beta$, calculated from the rest-frame FUV and NUV bands, does increase in value with inclination, this transition is minor compared to its spread.

\section{Analysis and Discussion} \label{sec:AnalysisDiscuss}

\subsection{Inclination Dependence of \texorpdfstring{$a_{\rm corr}$}{acorr} in Hybrid SFR Estimators} \label{sec:acorr}

\subsubsection{Influence of Inclination and SFH} \label{sec:acorrProp}

\begin{figure*}[t!]
\centerline{
\includegraphics[width=18cm]{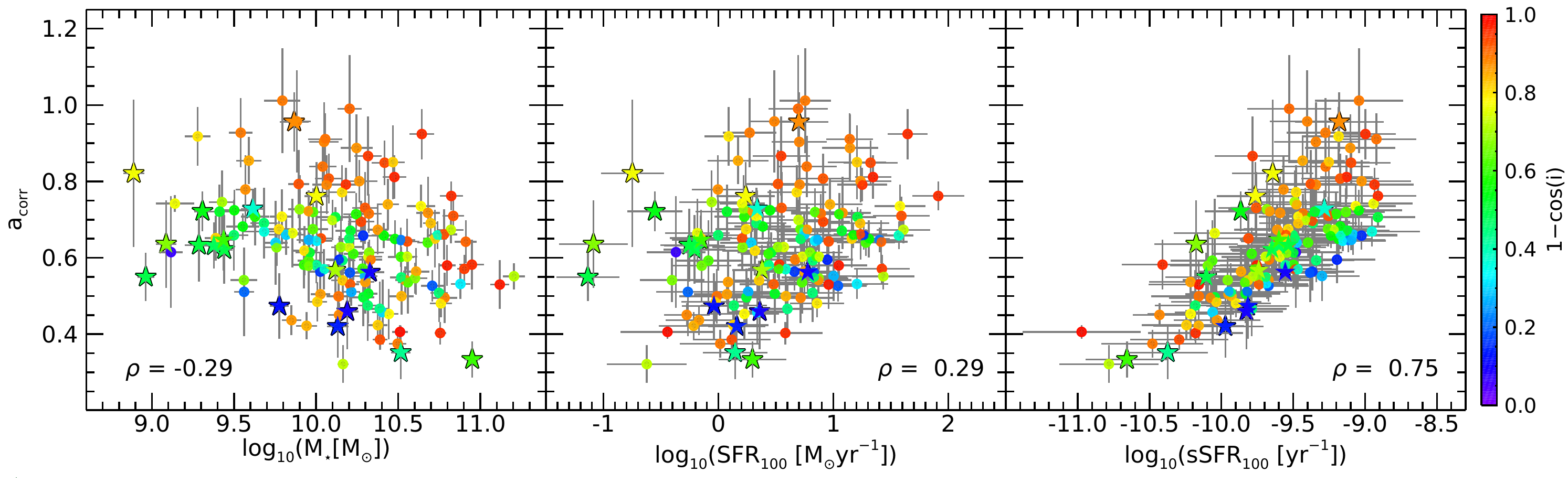}
}
\caption{
Each panel shows $a_{\rm corr}$ vs. a different physical property for the galaxies in our sample. Each galaxy is colored based on its median inclination as derived by \texttt{Lightning}. The circles are the CANDELS sample of galaxies, while the stars are the SINGS/KINGFISH sample. The Spearman correlation coefficient of each property vs. $a_{\rm corr}$ is shown in a lower corner of each panel. (Left) $a_{\rm corr}$ vs. total stellar mass ($M_\star$). The slight negative trend indicates that larger galaxies, which may have larger older populations, tend to have smaller values of $a_{\rm corr}$, with no clear trend with inclination. (Middle) $a_{\rm corr}$ vs. SFR averaged over the last 100 Myr ($\rm{SFR}_{100}$). The slight positive trend indicates that galaxies with younger populations tend to have larger values of $a_{\rm corr}$, with no clear trend with inclination. (Right) $a_{\rm corr}$ vs. sSFR averaged over the last 100 Myr ($\rm{sSFR}_{100}$). For a fixed $\rm{sSFR}_{100}$, galaxies that are more inclined typically have larger values of $a_{\rm corr}$.
}
\label{fig:acorrvprop}
\end{figure*}

Besides being dependent on inclination and other attenuation properties, the value of $a_{\rm corr}$ for a given galaxy is also dependent on the underlying stellar population or SFH \citep{2021arXiv211004314L}. While the FUV emission primarily samples young massive stars with stellar lifetimes $< 100$~Myr, the $L_{\rm TIR}$ samples the entire radiation field that is absorbed by dust, which is generated by stars of all stellar ages. Therefore, based on Equation~\ref{eq:acorr}, if we were to fix the attenuation and the luminosity of the young population (the FUV emission) while increasing the luminosity of the old population (the optical--to--near-IR, NIR, emission), we would expect $a_{\rm corr}$ to decrease in response, since $L_{\rm TIR}$ can be significantly impacted by the old stellar population \citep{2009ApJ...703.1672K}. Alternatively, if the $L_{\rm TIR}$ was fixed instead, we would expect $a_{\rm corr}$ to increase with an increase in the young FUV emitting population.

These trends with $a_{\rm corr}$ for our sample of galaxies can be seen in Figure~\ref{fig:acorrvprop}, which shows $a_{\rm corr}$ versus the total stellar mass ($M_\star$), the SFR averaged over the last 100 Myr ($\rm{SFR}_{100}$), and the sSFR averaged over the last 100 Myr (${\rm sSFR}_{100} \equiv {\rm SFR}_{100}/M_\star$). The total stellar mass is typically dominated by old stars, and $a_{\rm corr}$ can be seen to generally decrease with increasing $M_\star$, with a Spearman correlation coefficient of $\rho=-0.29$. As for $\rm{SFR}_{100}$, which is dominated by the young population, $a_{\rm corr}$ can be seen to generally increase with increasing $\rm{SFR}_{100}$ ($\rho=0.29$). However, these trends are both relatively weak, since $M_\star$ and $\rm{SFR}_{100}$ are usually highly correlated. A better measure of the underlying stellar population, besides the SFH itself, would be the $\rm{sSFR}_{100}$. Its trend with $a_{\rm corr}$ can be seen to be strong ($\rho=0.75$) and highly significant ($p$-value $< 10^{-25}$). 

This same trend between $a_{\rm corr}$ and $\rm{sSFR}_{100}$, ignoring inclination, was also found in several previous studies \citep[e.g.,][]{2014ApJ...795...89E,2017ApJ...851...10E,2016A&A...591A...6B,2021arXiv211004314L}. Notably, \citet{2016A&A...591A...6B} found a similarly strong trend in their sample of eight galaxies from KINGFISH. However, their sample was selected to exclude highly inclined galaxies ($1-\cos i <0.5$), which minimizes the inclination-dependent attenuation effects on $a_{\rm corr}$ seen in Figure~\ref{fig:incvacorr}. As can be seen in the right panel of Figure~\ref{fig:acorrvprop}, $a_{\rm corr}$ typically takes on a larger value as inclination increases for a fixed $\rm{sSFR}_{100}$. Therefore, any parameterization of $a_{\rm corr}$ must depend on both inclination and the $\rm{sSFR}_{100}$.

\begin{figure*}[t!]
\centerline{
\includegraphics[width=18cm]{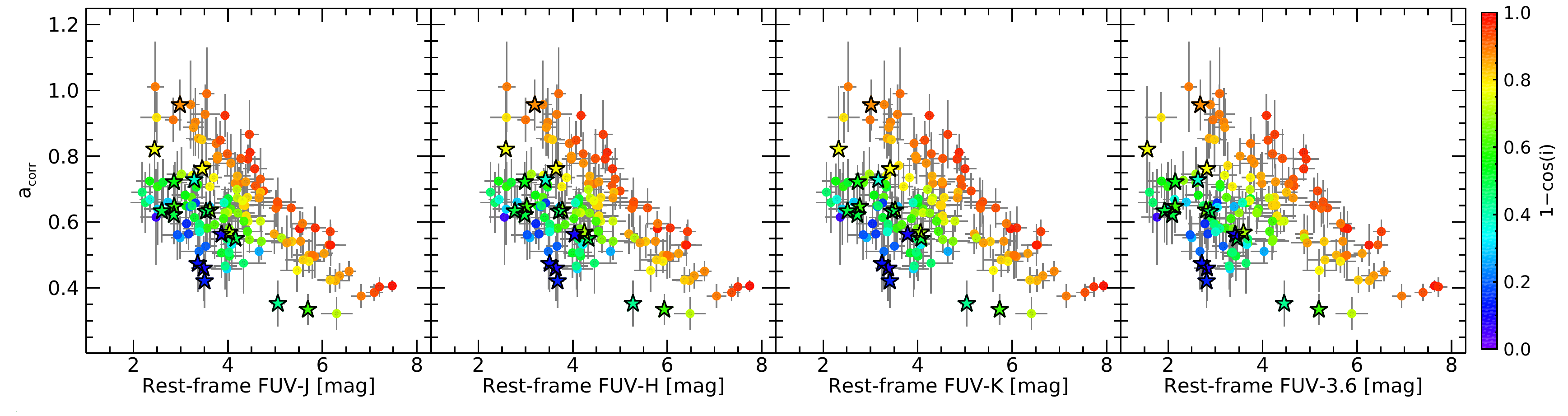}
}
\caption{
Each panel shows $a_{\rm corr}$ vs. a rest-frame model FUV--NIR color (FUV--$J$, FUV--$H$, FUV--$K$, FUV--3.6 $\mu$m from left to right) for the galaxies in our sample. Each galaxy is colored based on its median inclination as derived by \texttt{Lightning}. The circles are the CANDELS sample of galaxies, while the stars are the SINGS/KINGFISH sample. In all panels, a clear stratification can be seen in $a_{\rm corr}$--color space for galaxies of different inclinations.
}
\label{fig:acorrvcolor}
\end{figure*}
\begin{figure*}[t!]
\centerline{
\includegraphics[width=18cm]{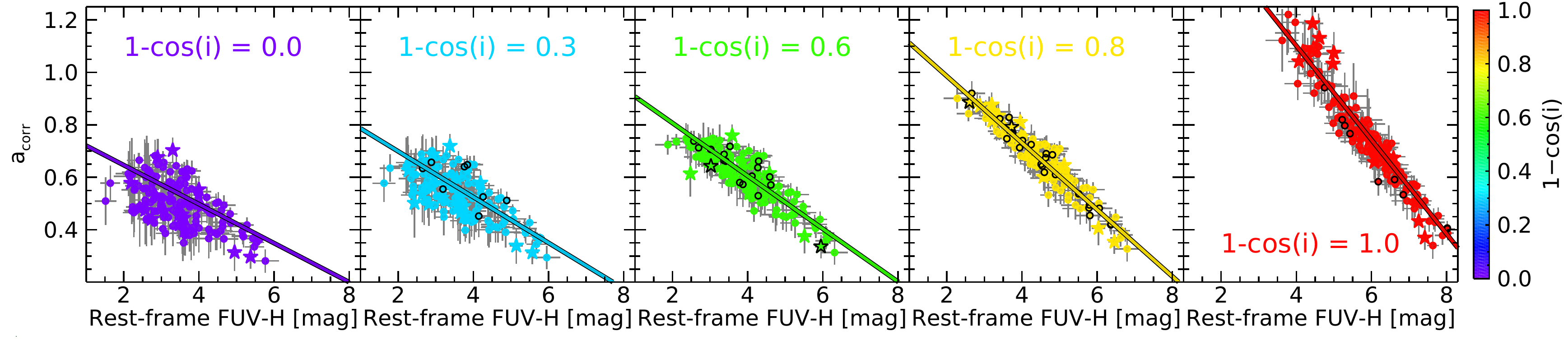}
}
\caption{
Each panel shows $a_{\rm corr}$ vs. rest-frame FUV--$H$ color for our simulated data for a span of inclination grid points, with the data in each panel being colored based on their inclination grid value ($1-\cos i = [0.0, 0.3, 0.6, 0.8, 1.0]$, from left to right). The circles are the CANDELS sample of galaxies, while the stars are the SINGS/KINGFISH sample. Points outlined in black indicate galaxies whose measured inclinations, in terms of $1-\cos i$, are within $\pm 0.05$ of the grid value. Each panel can be considered how the sample would appear if all galaxies were viewed from the respective inclination. The best-fit linear relation to the simulated data is shown in each panel. As inclination is increased from face-on to edge-on, the slope and intercept of the best-fit linear relations can be seen to decrease and increase, respectively.
}
\label{fig:acorrcolorperinc}
\end{figure*}

As noted in \citet{2016A&A...591A...6B}, a parameterization of $a_{\rm corr}$ with $\rm{sSFR}_{100}$ would not be a practical solution, as $\rm{sSFR}_{100}$ is a derived physical property rather than an observed quantity. Therefore, we utilized rest-frame FUV--NIR colors as in \citet{2016A&A...591A...6B} instead of $\rm{sSFR}_{100}$, since FUV--NIR colors are observable quantities and have been shown to be good tracers of $\rm{sSFR}_{100}$ \citep{2005ApJ...619L..39S,2016A&A...591A...6B}. Figure~\ref{fig:acorrvcolor} shows $a_{\rm corr}$ versus the rest-frame model FUV--$J$, FUV--$K$, FUV--$H$, and FUV--3.6 $\mu$m colors for the galaxies in our sample, where $J$, $H$, and $K$ are the 2MASS $J$, $H$, and $Ks$ bandpasses, and 3.6 $\mu$m is the Spitzer/IRAC 3.6 $\mu$m bandpass. In each panel of the figure, a clear stratification can be seen in the $a_{\rm corr}$--color space, where high-inclination galaxies ($1-\cos i \gtrsim 0.6$) populate regions of higher $a_{\rm corr}$ and FUV--NIR color compared to low-inclination galaxies ($1-\cos i \lesssim 0.6$). This striking trend can also be seen clearly in the simulated data in the left panel of Figure~\ref{fig:simdata}. In both the simulated data and Figure~\ref{fig:acorrvcolor}, the stratification of $a_{\rm corr}$ and FUV--NIR color with inclination is more pronounced at higher inclinations compared to lower inclinations due to the attenuation effects of inclination becoming more significant for inclinations of $1-\cos i \gtrsim 0.6$ \citep{2013MNRAS.432.2061C,2021ApJ...923...26D,2021ApJ...922L..32Z}.

\subsubsection{Relation between \texorpdfstring{$a_{\rm corr}$}{acorr} and Inclination} \label{sec:acorrRel}

Following the observed trends in Figure~\ref{fig:acorrvcolor}, we parameterized $a_{\rm corr}$ as a linear function of rest-frame FUV--NIR color for a given inclination using the functional form of
\begin{equation}
a_{corr}=b+m\times (\rm{FUV}-\rm{NIR}),
\label{eq:acorrcolor}
\end{equation}
where the linear coefficients $b$ and $m$ are both functions of inclination and unique to each FUV--NIR color. To derive these coefficients, we utilized our simulated data distributions described in Section~\ref{sec:SimData}, since using the data shown in Figure~\ref{fig:acorrvcolor} would result in a sparse population of inclination--$a_{\rm corr}$--color space. The simulated data increased the amount of data at each inclination, since each galaxy was simulated for a grid of viewing angles.

\begin{figure*}[t!]
\centerline{
\includegraphics[width=18cm]{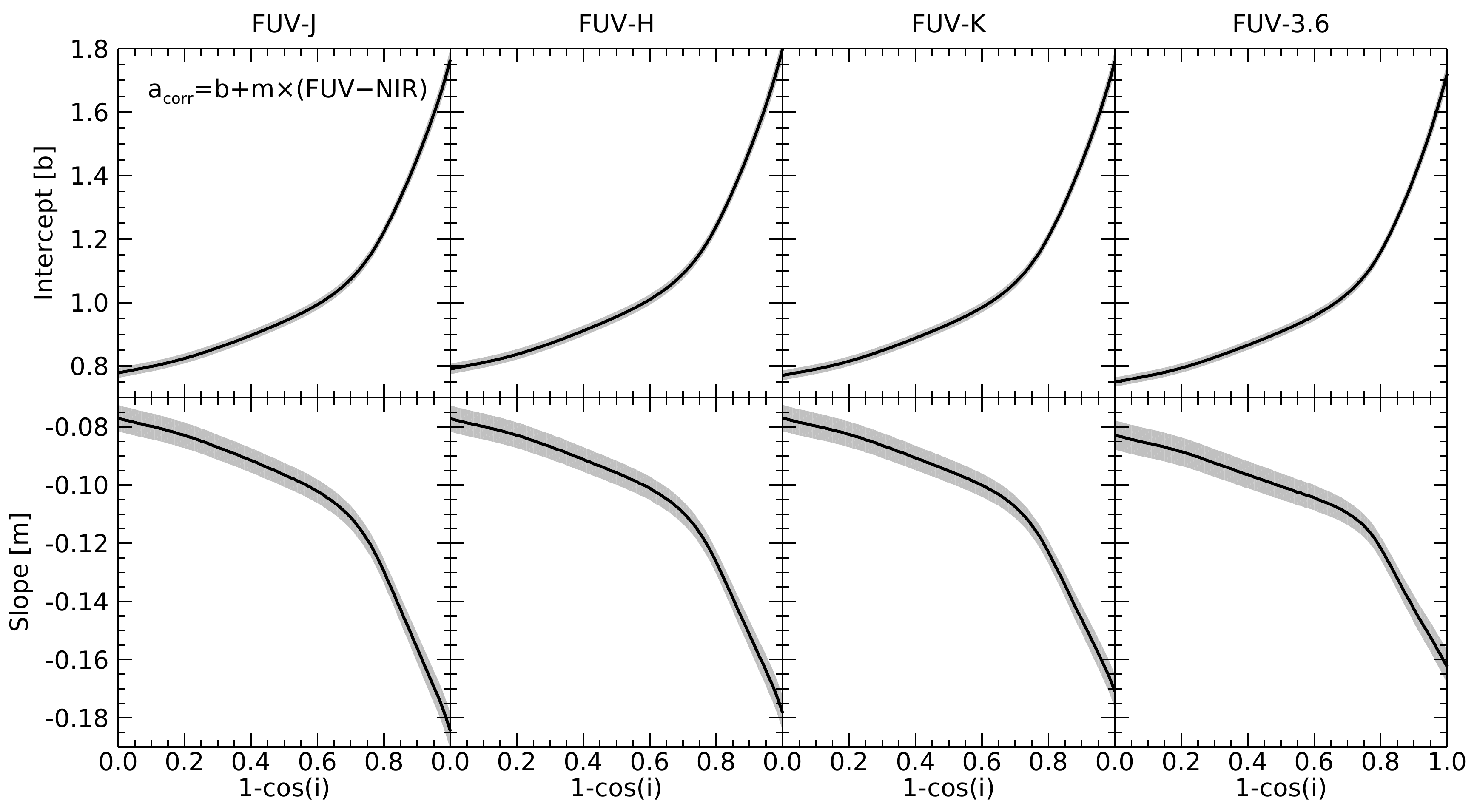}
}
\caption{
Linear coefficients for Equation~\ref{eq:acorrcolor} vs. inclination for the four rest-frame FUV--NIR colors. The black line shows the derived values at each inclination, with the gray shaded region giving the derived uncertainties.
}
\label{fig:colorcoeff}
\end{figure*}

For each inclination grid point of the simulated data, we used the median of the distributions of $a_{\rm corr}$ and FUV--NIR color of each galaxy (e.g., the solid black line in the left panel of Figure~\ref{fig:simdata}) as data points and fitted the linear relationship of Equation~\ref{eq:acorrcolor} to these median values. The corresponding standard deviations of the $a_{\rm corr}$ and FUV--NIR color distributions were included as uncertainties during the fitting process. The fitting was repeated for each inclination grid point, resulting in derived $b$ and $m$ values with corresponding uncertainties at each of the inclination grid points. An example of this process can be seen in Figure~\ref{fig:acorrcolorperinc}, which shows the simulated data and best-fit $a_{\rm corr}$ versus FUV--NIR color relation at various inclination grid points. From the figure, the slope and intercept of the linear relation can be seen to decrease and increase with inclination, respectively. These resulting trends in $b$ and $m$ versus inclination can be more clearly seen in Figure~\ref{fig:colorcoeff} for each FUV--NIR color. For each color, the linear coefficients show very similar trends, with more rapid changes in value occurring at high inclinations ($1-\cos i > 0.7$), where the attenuation effects of inclination become more significant.

\begin{deluxetable*}{cccccc}
\tablecaption{Polynomial Coefficients to estimate $a_{\rm corr}$ as a function of inclination and \textbf{rest-frame} FUV--NIR color via Equation~\ref{eq:acorrcolorinc}. \label{table:colorcoeff}}
\tablecolumns{6}
\tablehead{  & \multicolumn{5}{c}{Polynomial Coefficients for Intercept $b$} \\Color & $b_0$ & $b_1$ & $b_2$ & $b_3$ & $b_4$}
\startdata
FUV--$J$ &   \phm{$-$}$ 0.7820\pm0.0075$ & \phm{$-$}$ 0.0298\pm0.1090$ & $ 1.4679\pm0.4645$ & $-3.1348\pm0.7284$ & $ 2.6395\pm0.3762$ \\
FUV--$H$ &   \phm{$-$}$ 0.7950\pm0.0078$ & \phm{$-$}$ 0.0081\pm0.1137$ & $ 1.6177\pm0.4840$ & $-3.4210\pm0.7578$ & $ 2.8188\pm0.3908$ \\
FUV--$K$ &   \phm{$-$}$ 0.7759\pm0.0073$ & $-0.0248\pm0.1065$ & $ 1.7531\pm0.4544$ & $-3.6430\pm0.7130$ & $ 2.9165\pm0.3684$ \\
FUV--3.6 & \phm{$-$}$ 0.7579\pm0.0070$ & $-0.1099\pm0.1022$ & $ 2.2370\pm0.4365$ & $-4.5584\pm0.6857$ & $ 3.4086\pm0.3548$ \\
\tableline
 & \multicolumn{5}{c}{Polynomial Coefficients for Slope $m$} \\
Color & $m_0$ & $m_1$ & $m_2$ & $m_3$ & \\
\tableline
FUV--$J$ &   $-0.0741\pm0.0017$ & $-0.0819\pm0.0149$ & $ 0.1931\pm0.0351$ & $-0.2230\pm0.0235$ &   \\
FUV--$H$ &   $-0.0743\pm0.0017$ & $-0.0797\pm0.0149$ & $ 0.1865\pm0.0349$ & $-0.2118\pm0.0232$ &   \\
FUV--$K$ &   $-0.0742\pm0.0017$ & $-0.0774\pm0.0148$ & $ 0.1770\pm0.0345$ & $-0.1974\pm0.0229$ &   \\
FUV--3.6 & $-0.0797\pm0.0019$ & $-0.0832\pm0.0161$ & $ 0.1835\pm0.0371$ & $-0.1847\pm0.0244$ &   \\
\enddata
\end{deluxetable*}

To account for the variation in $b$ and $m$ with inclination, we fitted polynomials to the derived $b$ and $m$ values utilizing their corresponding uncertainties. The degree of the polynomial was selected by minimizing the Akaike information criterion (AIC). For all FUV--NIR colors, this resulted in fourth- and third-order polynomials being chosen for the $b$ and $m$ parameters, respectively. Incorporating this inclination dependence on $b$ and $m$, Equation~\ref{eq:acorrcolor} can be rewritten as
\begin{equation}
\begin{aligned}
a_{\rm corr} = & \sum_{n=0}^{4} b_n (1-\cos i)^n +\\ & \sum_{n=0}^{3} m_n (1-\cos i)^n \times (\rm{FUV}-\rm{NIR}),
\end{aligned}
\label{eq:acorrcolorinc}
\end{equation}
where $b_n$ and $m_n$ are the polynomial coefficients of $b$ and $m$, which can be found in Table~\ref{table:colorcoeff} along with their corresponding uncertainty for each FUV--NIR color. Therefore, Equation~\ref{eq:acorrcolorinc} gives a parametric estimation of $a_{\rm corr}$ that only depends on the observable quantities of FUV--NIR color and inclination, allowing for an easy-to-use inclination-dependent hybrid SFR estimator.

\begin{figure*}[t!]
\centerline{
\includegraphics[width=18cm]{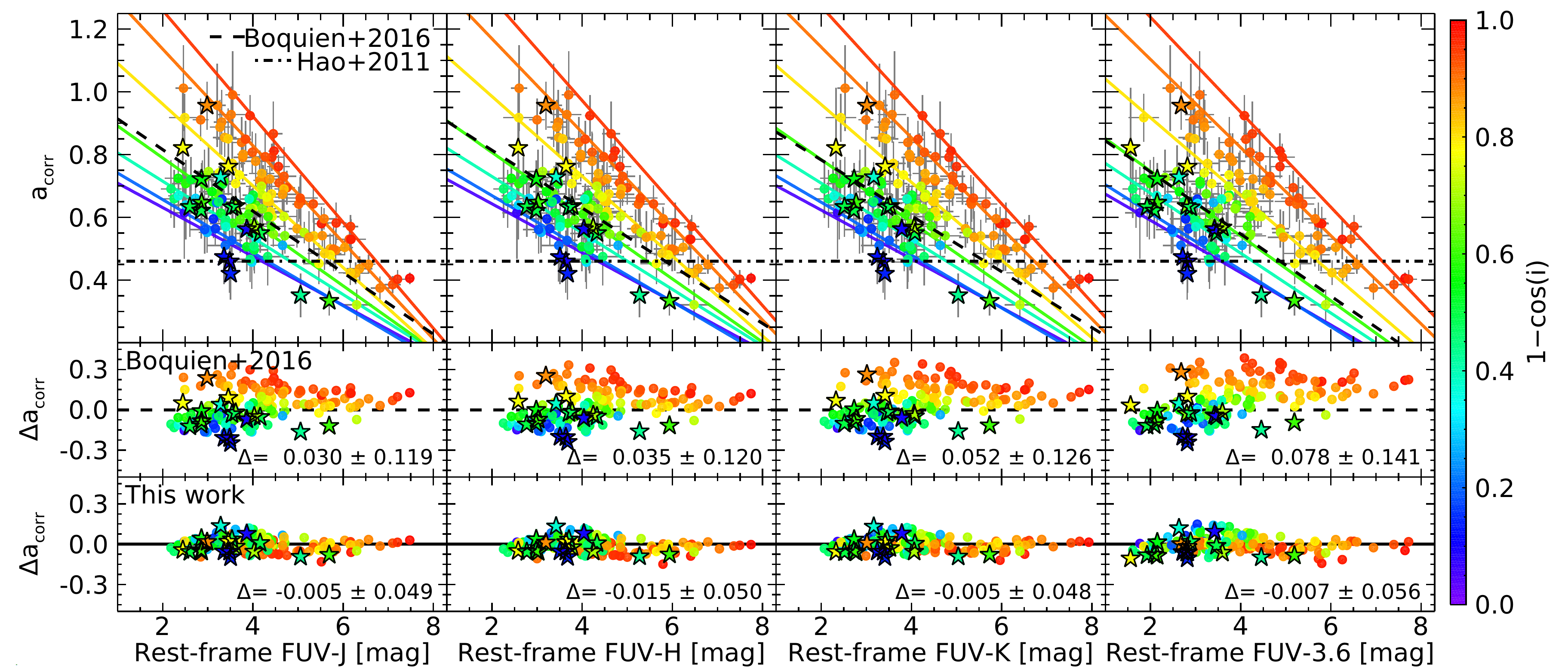}
}
\caption{
In all panels, the circles are the CANDELS sample of galaxies, while the stars are the SINGS/KINGFISH sample. Each galaxy is colored based on its median inclination as derived by \texttt{Lightning}. (Upper row) Each panel shows $a_{\rm corr}$ vs. a rest-frame model FUV--NIR color (FUV--$J$, FUV--$H$, FUV--$K$, FUV--3.6 $\mu$m from left to right) for the galaxies in our sample. The parametric estimation of $a_{\rm corr}$ from this study is shown as the solid colored lines, with the color indicating the inclination used in the calculation ($1-\cos i = [0.05, 0.2, 0.4, 0.6, 0.8, 0.9, 0.95]$). The dashed-dotted and dashed lines are the $a_{\rm corr}$ value from \citet{2011ApJ...741..124H} and $a_{\rm corr}$--color relation from \citet{2016A&A...591A...6B}, respectively, for the FUV and $L_{\rm TIR}$. (Middle row) Difference between $a_{\rm corr}$ derived from \texttt{Lightning} and $a_{\rm corr}$ derived from the \citet{2016A&A...591A...6B} relation vs. an FUV--NIR color. The delta in the lower right corner is the mean and standard deviation of $\Delta a_{\rm corr}$ (i.e., the mean and scatter of the residuals).  (Lower row) The difference between $a_{\rm corr}$ derived from \texttt{Lightning} and $a_{\rm corr}$ derived from the parametric relation in this work vs. an FUV--NIR color. The delta in the lower right corner is the mean and standard deviation of $\Delta a_{\rm corr}$.
}
\label{fig:acorrvcolorfit}
\end{figure*}

 \subsubsection{Comparison with Past Studies} \label{sec:acorrCompare}
 
The parametric estimation of $a_{\rm corr}$ as a function of inclination and rest-frame FUV--NIR color can be seen in the upper row of Figure~\ref{fig:acorrvcolorfit}. This upper row is the same as Figure~\ref{fig:acorrvcolor}, but it now includes the parametric estimation of $a_{\rm corr}$ from Equation~\ref{eq:acorrcolorinc} as the solid colored lines, with the color indicating the inclination used in the calculation. Additionally, the corresponding $a_{\rm corr}$ value from \citet{2011ApJ...741..124H} and $a_{\rm corr}$--color relation from \citet{2016A&A...591A...6B} for the FUV and $L_{\rm TIR}$ are shown as the dashed-dotted and dashed lines, respectively. From this upper row, it can be seen that the value of $a_{\rm corr}$ from \citet{2011ApJ...741..124H} is much lower than the derived $a_{\rm corr}$ values for the vast majority of our galaxies. This discrepancy is caused by the differences in the utilized galaxy samples. \citet{2011ApJ...741..124H} used a sample of galaxies including both late- and early-type galaxies, where we selected only late-type, star-forming galaxies. Therefore, our sample will, on average, have galaxies with higher sSFR, which will correspondingly result in larger values of $a_{\rm corr}$.

As for the \citet{2016A&A...591A...6B} $a_{\rm corr}$--color relation, the upper row of panels shows near agreement with our parameterization for $1-\cos i \approx 0.6$ ($i \approx 66^\circ$). This coinciding inclination supports our methodology, since the majority of the \citet{2016A&A...591A...6B} galaxy sample had $i = 50^\circ$--$60^\circ$. In the bottom two rows of Figure~\ref{fig:acorrvcolorfit}, we show residuals of $a_{\rm corr}$ ($\Delta a_{\rm corr}$; the difference between $a_{\rm corr}$ derived from \texttt{Lightning} and $a_{\rm corr}$ derived from the \citealt{2016A&A...591A...6B} relation or the parametric relation in this work) versus FUV--NIR color. From these panels, it can be seen that the \citet{2016A&A...591A...6B} relation, on average, is consistent with our data but results in large scatter that has a clear inclination dependence, with more face-on galaxies typically having their $a_{\rm corr}$ overestimated and more edge-on galaxies having their $a_{\rm corr}$ underestimated. However, the parameterization in this work results in residuals that have a scatter that is less than half that from the \citet{2016A&A...591A...6B} relation and no inclination dependence, implying that the effects of inclination are being properly accounted for in our relation. Therefore, our parameterization is the first, to our knowledge, that accounts for both the effects of SFH and inclination that are expected to be present when determining $a_{\rm corr}$. We note, however, that the $a_{\rm corr}$ relation presented above has a specific range of applicability and a few caveats, which are discussed in Section~\ref{sec:Applic}.

\subsection{Inclination Dependence of the \texorpdfstring{$A_{\rm FUV}$}{AFUV}--\texorpdfstring{$\beta$}{Beta} Relation} \label{sec:M99}

\subsubsection{Influence of Inclination and SFH} \label{sec:M99Prop}

\begin{figure*}[t!]
\centerline{
\includegraphics[width=18cm]{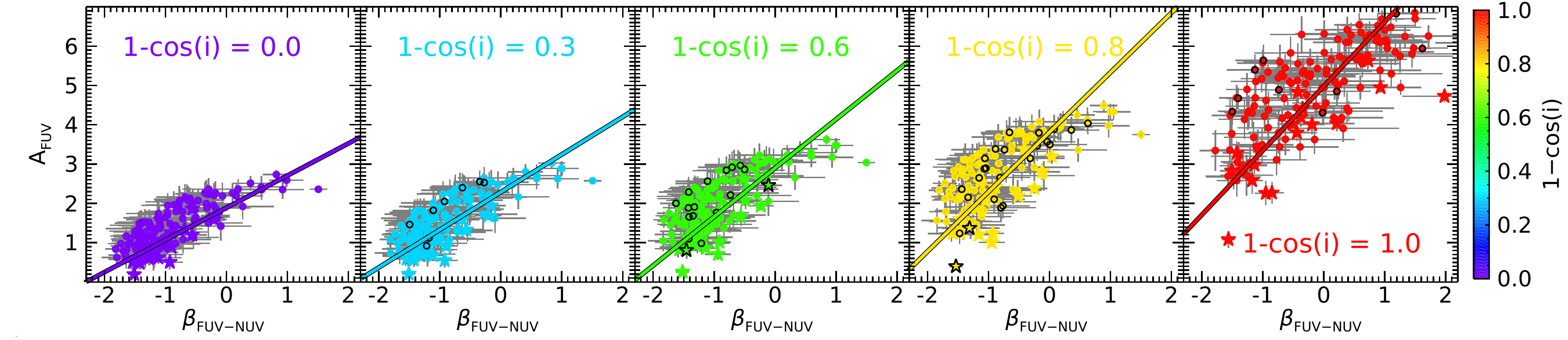}
}
\caption{
Each panel shows $A_{\rm FUV}$ vs. $\beta$, calculated from the rest-frame model FUV and NUV bands ($\beta_{\rm FUV-NUV}$), for our simulated data for a span of inclination grid points, with the data in each panel being colored based on their inclination grid value ($1-\cos i = [0.0, 0.3, 0.6, 0.8, 1.0]$, from left to right). The circles are the CANDELS sample of galaxies, while the stars are the SINGS/KINGFISH sample. Points outlined in black indicate galaxies whose measured inclinations, in terms of $1-\cos i$, are within $\pm 0.05$ of the grid value. Each panel can be considered how the sample would appear if all galaxies were viewed from the respective inclination. The best-fit linear relation to the simulated data is shown in each panel. As inclination is increased from face-on to edge-on, the slope of the best-fit linear relation can also be seen to increase, while the $\beta$-intercept only decreases at the largest inclinations.
}
\label{fig:afuvbetaperinc}
\end{figure*}

Based on the definition of the $A_{\rm FUV}$--$\beta$ relation used in Equation~\ref{eq:AUVbeta}, the calibrated parameter $a_{\beta}$ should solely depend on the choice of attenuation curve, and $\beta_0$ should only depend on the SFH of the galaxy, since we assumed a fixed metallicity and IMF. In our study, we chose to use inclination-dependent attenuation curves, which depend on three free parameters: $\tau_B^f$ (the face-on optical depth in the $B$ band), $F$ (the galaxy clumpiness factor), and inclination. While inclination is a quantity that can be readily determined from basic observations, $\tau_B^f$ and $F$ are intrinsic properties that can only be derived from modeling. Therefore, our parameterization of $a_\beta$ can only be a function of inclination, since it is the only observable property, and any scatter in the parameterization will be due to the variation in other attenuation parameters at a given inclination.

As for $\beta_0$, in theory, its value will be unique for each galaxy, since it is dependent on the SFH. However, in application, a fixed value of $\beta_0$ for a sample of galaxies is generally utilized \citep[e.g.,][]{1999ApJ...521...64M,2011ApJ...726L...7O,2012A&A...539A.145B,2018ApJ...869..161W}, since the SFH of a galaxy is not an observable property. While the SFH could be approximated using a rest-frame FUV--NIR or a comparable color, the $A_{\rm FUV}$--$\beta$ relation is typically helpful when minimal UV observational data are available, preventing use of an SFH proxy. Therefore, we do not include any color dependence in our $A_{\rm FUV}$--$\beta$ relation and note that additional scatter and potential systematic effects will be present in the relation due to not incorporating any SFH dependence on $\beta_0$.

Finally, as discussed in Section~\ref{sec:PhysProp}, $a_\beta$ and $\beta_0$ will depend on the choice of UV bandpasses utilized in the calculation. While $\beta_0$ will have minimal variation from the choice of UV bandpasses due to it being a dust-free property, $a_\beta$ can be biased to larger values if a chosen UV bandpass is contaminated by the 2175 \AA\ bump feature. Therefore, in the next section, we derive two inclination-dependent $A_{\rm FUV}$--$\beta$ relations using the combination of bandpasses discussed in Section~\ref{sec:PhysProp}. The first uses the combination of the GALEX FUV and NUV bands, which will suffer from UV bump contamination. The second uses the combination of the GALEX FUV and HST WFC3/F275W bands, neither of which overlap the bump feature region.

\subsubsection{Inclination-dependent \texorpdfstring{$A_{\rm FUV}$}{AFUV}--\texorpdfstring{$\beta$}{Beta} Relation} \label{sec:M99Rel}

Since the relation between $A_{\rm FUV}$ and $\beta$ given in Equation~\ref{eq:AUVbeta} is linear, we followed the same method as in Section~\ref{sec:acorrRel} when deriving $a_\beta$ and $\beta_0$ for the $A_{\rm FUV}$--$\beta$ relations. This method again relied on our simulated data distributions at each inclination. For each inclination grid point of the simulated data, we utilized the median of the distributions of $A_{\rm FUV}$ and $\beta$ of each galaxy (e.g., the solid black line in the right panel of Figure~\ref{fig:simdata}) as data points and fitted the linear relationship of Equation~\ref{eq:AUVbeta} to these data. The corresponding standard deviations of the $A_{\rm FUV}$ and $\beta$ distributions were included as uncertainties during the fitting process. The fitting was repeated for each inclination grid point, resulting in derived $a_\beta$ and $\beta_0$ values with corresponding uncertainties at each of the inclination grid points. An example of the process can be seen in Figure~\ref{fig:afuvbetaperinc}, which shows the simulated data and best-fit relation at various inclination grid points.

\begin{figure*}[t!]
\centerline{
\includegraphics[width=18cm]{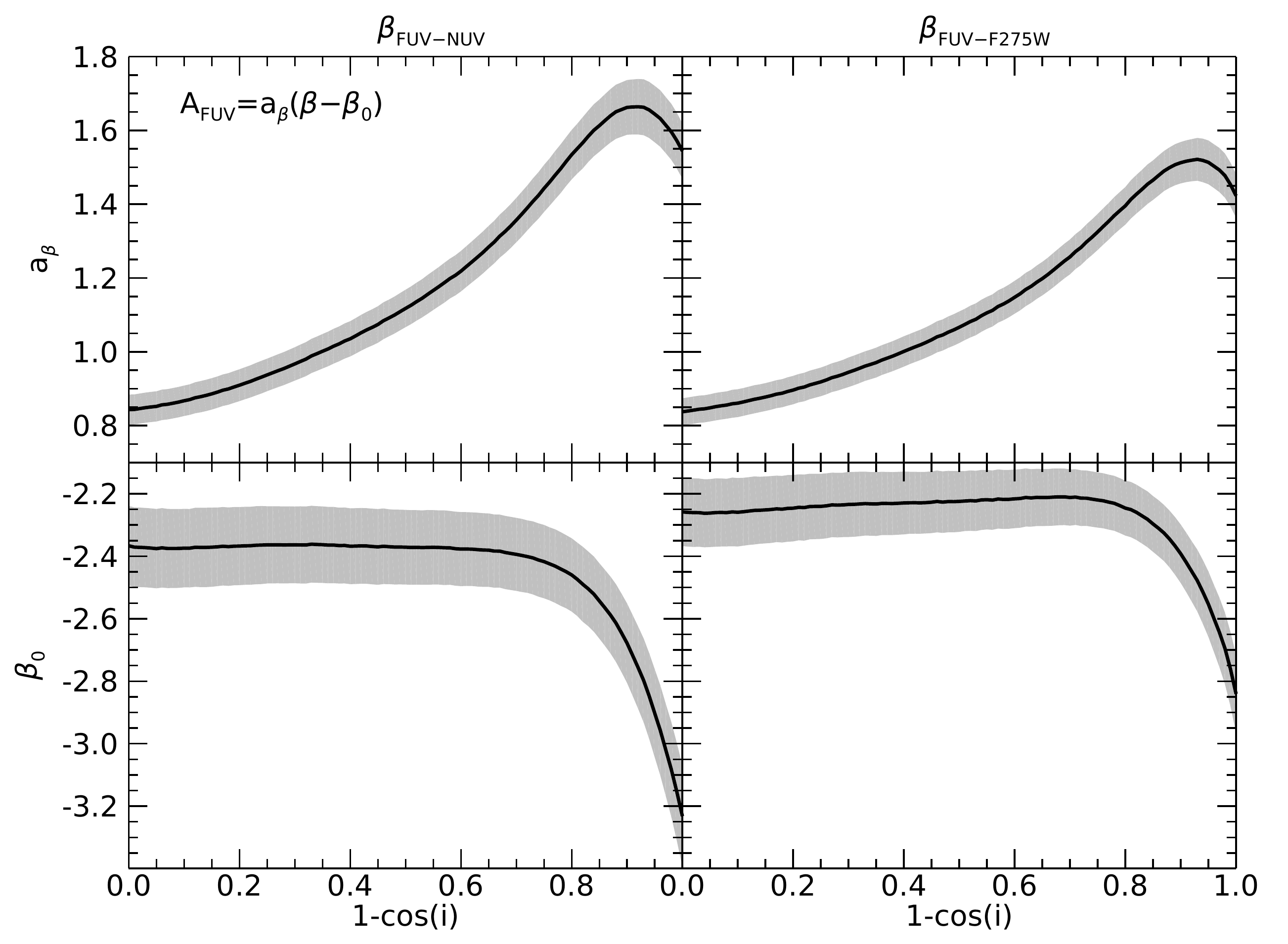}
}
\caption{
Linear coefficients, $a_\beta$ and $\beta_0$, for Equation~\ref{eq:AUVbeta} vs. inclination for the two combinations of UV bandpasses. The black line shows the derived values at each inclination, with the gray shaded region giving the derived uncertainties.
}
\label{fig:abetabeta0coeff}
\end{figure*}

The resulting trends in $a_\beta$ and $\beta_0$ versus inclination are shown in Figure~\ref{fig:abetabeta0coeff} for the two sets of UV bandpasses used when calculating $\beta$. For both sets of bandpasses, $a_\beta$ and $\beta_0$ show similar trends. As expected, $a_\beta$ increases in value as inclination increases from face-on to edge-on. However, above $1-\cos i \approx 0.9$, $a_\beta$ begins to decrease with increasing inclination. This decrease is correlated to the unexpected result of $\beta_0$ decreasing at $1-\cos i > 0.75$.  Theoretically, $\beta_0$ is expected to be inclination-independent, since it is a dust-free property. Therefore, it should be constant as a function of inclination, and the observed decrease at high inclinations could be due to our various simplifying assumptions. For example, the SFH dependence of $\beta_0$ could be disguised as an inclination dependence at these high inclinations. Additionally, the assumption in the $A_{\rm FUV}$--$\beta$ relation that the UV slope is linearly related to UV attenuation could be too simplified for high-inclination galaxies.

Rather than attempting to correct for these simplifying assumptions (i.e., adding an SFH dependence, changing from a linear relation, etc.), we only add an inclination dependence to $a_\beta$ and $\beta_0$ to maintain the $A_{\rm FUV}$--$\beta$ relation's simplistic format. To account for the variation in $a_\beta$ and $\beta_0$ with inclination for both sets of UV bandpasses, we fitted polynomials to the corresponding $a_\beta$ and $\beta_0$ values in Figure~\ref{fig:abetabeta0coeff} utilizing their derived uncertainties. We selected the degree of the polynomials by minimizing the AIC. For both sets of bandpasses, this resulted in fifth- and fourth-order polynomials being chosen for $a_\beta$ and $\beta_0$, respectively. Incorporating this inclination dependence on $a_\beta$ and $\beta_0$, Equation~\ref{eq:AUVbeta} can be rewritten as
\begin{equation}
\begin{aligned}
A_{\rm FUV} = & \sum_{n=0}^{5} a_{\beta, n} (1-\cos i)^n \times \\ & \Bigg(\beta-\sum_{n=0}^{4} \beta_{0, n} (1-\cos i)^n\Bigg)
\end{aligned}
\label{eq:afuvbetainc}
\end{equation}
where $a_{\beta, n}$ and $\beta_{0, n}$ are the polynomial coefficients of $a_\beta$ and $\beta_0$, which can be found in Table~\ref{table:abetacoeff} along with their corresponding uncertainty for each set of UV bandpasses.

%
\begin{deluxetable}{ccc}
\tablecaption{Polynomial Coefficients to Estimate $A_{\rm FUV}$ as a Function of $\beta$ and Inclination via Equation~\ref{eq:afuvbetainc}. \label{table:abetacoeff}}
\tablecolumns{3}
\tablehead{ &  UV Bump & No UV Bump \\Coefficients & FUV--NUV & FUV--F275W}
\startdata
$a_{\beta, 0}$ & \phm{$-$0} $  0.8564\pm \phn  0.0230$ & \phm{$-$0} $  0.8507\pm0.0206$ \\
$a_{\beta, 1}$ & \phm{0}    $ -0.4759\pm \phn  0.5065$ & \phm{0}    $ -0.3892\pm0.4447$ \\
$a_{\beta, 2}$ & \phm{$-$0} $  7.0243\pm \phn  3.3703$ & \phm{$-$0} $  5.8447\pm2.9072$ \\
$a_{\beta, 3}$ & \phm{}     $-21.4069\pm \phn  9.0246$ & \phm{}     $-17.6998\pm7.6714$ \\
$a_{\beta, 4}$ & \phm{$-$}  $ 29.5716\pm10.3862$ & \phm{$-$}  $ 24.0990\pm8.7243$ \\
$a_{\beta, 5}$ & \phm{}     $-14.0028\pm \phn  4.2785$ & \phm{}     $-11.2503\pm3.5597$ \\
\tableline
$\beta_{0, 0}$ & \phm{0}    $ -2.4084\pm \phn  0.0596$ & \phm{0}    $ -2.2972\pm 0.0508$ \\
$\beta_{0, 1}$ & \phm{$-$0} $  0.9974\pm \phn  0.8306$ & \phm{$-$0} $  0.9985\pm 0.6937$ \\
$\beta_{0, 2}$ & \phm{0}    $ -5.7388\pm \phn  3.4059$ & \phm{0}    $ -5.2784\pm 2.7990$ \\
$\beta_{0, 3}$ & \phm{$-$}  $ 11.5513\pm \phn  5.1544$ & \phm{$-$}  $ 10.4165\pm 4.1830$ \\
$\beta_{0, 4}$ & \phm{0}    $ -7.5682\pm \phn  2.5757$ & \phm{0}    $ -6.6026\pm 2.0700$ \\
\enddata
\end{deluxetable}

\subsubsection{Comparison with Past Studies} \label{sec:M99Compare}

\begin{figure*}[t!]
\centerline{
\includegraphics[width=18cm]{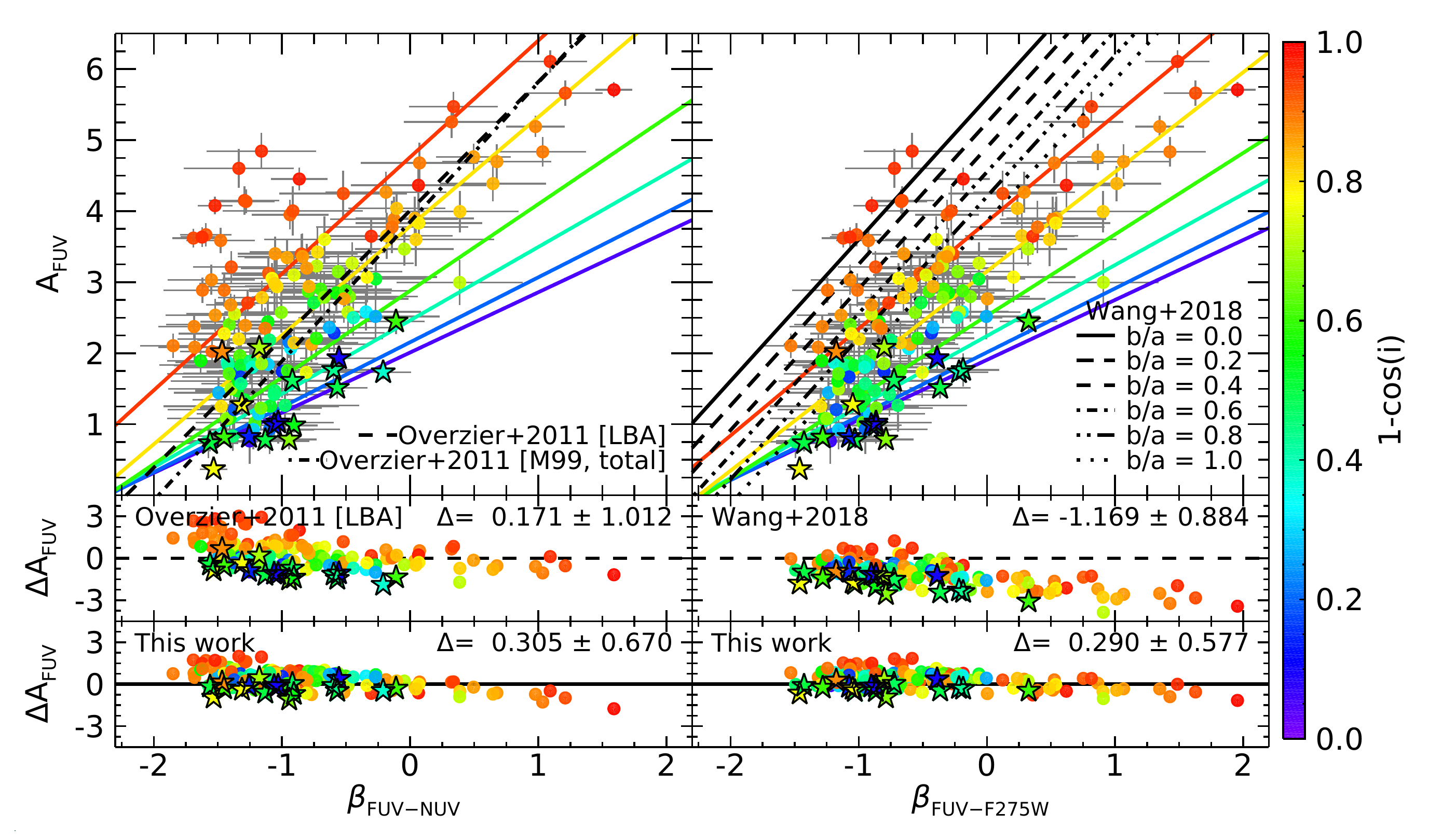}
}
\caption{
In all panels, the circles are the CANDELS sample of galaxies, while the stars are the SINGS/KINGFISH sample. Each galaxy is colored based on its median inclination as derived by \texttt{Lightning}. (Upper row) Each panel shows $A_{\rm FUV}$ vs. $\beta$ calculated using the rest-frame model FUV and NUV bands ($\beta_{\rm FUV-NUV}$) and FUV and F275W bands ($\beta_{\rm FUV-F275W}$) in the left and right panels, respectively. The corresponding inclination-dependent $A_{\rm FUV}$--$\beta$ relations from this study are shown as the solid colored lines, with the color indicating the inclination used in the calculation ($1-\cos i = [0.05, 0.2, 0.4, 0.6, 0.75]$). The dashed and dashed-dotted lines in the left panel are the relations from \citet{2011ApJ...726L...7O} for their LBA sample and the \citet{1999ApJ...521...64M} sample, respectively. The black lines of changing line style in the right panel are the inclination-dependent \citet{2018ApJ...869..161W} relation, where each line style represents a different value of axis ratio. (Middle row) Difference between $A_{\rm FUV}$ derived from \texttt{Lightning} and $A_{\rm FUV}$ derived from the \citet{2011ApJ...726L...7O} LBA relation and the \citet{2018ApJ...869..161W} relation utilizing each galaxies' measured axis ratio on the left and right, respectively. The delta in the upper right corner is the mean and standard deviation of $\Delta A_{\rm FUV}$ (i.e., the mean and scatter of the residuals). (Lower row) Difference between $A_{\rm FUV}$ derived from \texttt{Lightning} and $A_{\rm FUV}$ derived from the inclination-dependent $A_{\rm FUV}$--$\beta$ relations in this work. The delta in the upper right corner is the mean and standard deviation of $\Delta A_{\rm FUV}$.
}
\label{fig:afuvvbetafit}
\end{figure*}

The inclination-dependent $A_{\rm FUV}$--$\beta$ relations for each set of UV bandpasses are shown in the upper row of Figure~\ref{fig:afuvvbetafit}. This upper row is the same as Figure~\ref{eq:AUVbeta} but includes these inclination-dependent relations as the solid colored lines, with the color indicating the inclination used in the calculation. Additionally, we show different $A_{\rm FUV}$--$\beta$ relations derived in past studies.

In the left column, we compare our results with the two relations derived in \citet{2011ApJ...726L...7O}: one from their sample of Lyman break analogs (LBAs) and the other from the same sample of galaxies in \citet{1999ApJ...521...64M}. These relations were calibrated using the IRX--$\beta$ relation, where the $\beta$ values were calculated using the GALEX FUV and NUV bands, which will share the same bias as our inclination-dependent relation calculated using these bands. We find that the LBA sample relation has a similar $\beta_0$ value ($\beta_0 = -2.22$) as that of our relation at low-to-moderate inclinations ($\beta_0 \approx -2.35$), while the \citet{1999ApJ...521...64M} sample relation is significantly higher ($\beta_0 = -1.96$). Therefore, in the middle panel of the left column, we show residuals of $A_{\rm FUV}$ ($\Delta A_{\rm FUV}$; the difference between $A_{\rm FUV}$ derived from \texttt{Lightning} and $A_{\rm FUV}$ derived from the LBA relation) versus $\beta$ for the LBA relation. From this panel, it can be seen that the LBA relation from \citet{2011ApJ...726L...7O} has a clear inclination dependence in the residuals, with low-inclination galaxies typically having their $A_{\rm FUV}$ overestimated and high-inclination galaxies typically having theirs underestimated. However, the relation in our work results in residuals (lower left panel of Figure~\ref{fig:afuvvbetafit}) with minimal inclination dependence. Also, the scatter in the residuals of our relation is smaller than the residuals of the LBA relation by a factor of $\approx$1.5, indicating that its inclination dependence is accounting for some additional variation present in the $A_{\rm FUV}$--$\beta$ relation.

In the right column of Figure~\ref{fig:afuvvbetafit}, we compare our results to the inclination-dependent  $A_{\rm FUV}$--$\beta$ relation from \citet{2018ApJ...869..161W}, which utilized axis ratio ($q=b/a$; $q=0$ is edge-on and $q=1$ is face-on) rather than inclination. To briefly explain the derivation of this relation, its inclination dependence was derived by first assuming the hybrid SFR estimators are inclination-independent, and then using this assumption to correct the $A_{\rm FUV}$--$\beta$ relation for inclination. This inclination correction was then added to the $\beta_0$ term, while $a_\beta$ was fixed to a constant value. Also, the $\beta$ values used in the derivation were calculated by fitting a power law to three observed UV photometric data points, all of which were selected to avoid the UV bump feature. Therefore, we compared this relation to our relation calculated using the FUV and F275W bands, since both relations should avoid the bias introduced by the presence of the UV bump. 

The upper right panel of Figure~\ref{fig:afuvvbetafit} shows the inclination-dependent \citet{2018ApJ...869..161W} relation as the black lines of changing line style, where each line style represents a different value of axis ratio. From this panel, it can be seen that the \citet{2018ApJ...869..161W} relation overestimates $A_{\rm FUV}$ for practically all of the galaxies in our sample. This is clearly seen in the residuals for the \citet{2018ApJ...869..161W} relation shown in the middle panel, where the $A_{\rm FUV}$ values from the \citet{2018ApJ...869..161W} relation were calculated utilizing the axis ratios of our galaxies as described in Section~\ref{sec:Data}. The reason for this overestimation by the \citet{2018ApJ...869..161W} relation for our sample comes from their critical assumption that hybrid SFR estimators are inclination-independent, which this paper has shown to not be the case. Ignoring this inclination dependence in their calculation is causing overestimates of $A_{\rm FUV}$, especially at low inclinations, where the hybrid SFR estimator is likely overestimating the SFR.

\subsection{Range of Applicability and Caveats} \label{sec:Applic}

It is important to stress that the relations for unattenuating the FUV luminosity presented in this paper were derived from a specific sample of disk-dominated galaxies (see Section~\ref{sec:Data}). Therefore, their use should be limited to galaxies whose physical properties fall within the range of our sample. Extrapolating their use to galaxies outside this range could result in unrealistic unattenuated luminosities. For the inclination- and color-dependent hybrid SFR estimator, the rest-frame FUV--NIR colors should be within the following ranges:
\begin{itemize}
\item[] $2.18<$ FUV--$J$ $<7.48$ mag,
\item[] $2.26<$ FUV--$H$ $<7.74$ mag,
\item[] $2.07<$ FUV--$K$ $<7.93$ mag, and
\item[] $1.56<$ FUV--3.6 $<7.72$ mag.
\end{itemize}
As for the inclination-dependent $A_{\rm FUV}$--$\beta$ relation, $\beta$ values should fall within
\begin{itemize}
\item[] $-1.85<\beta_{\rm bump}<1.59$ and
\item[] $-1.53<\beta_{\rm no\ bump}<1.96$
\end{itemize}
for galaxies that have and do not have UV bump-contaminated observations, respectively. Additionally, galaxies, as per Section~\ref{sec:Data}, should be star-forming disk galaxies with a minimal bulge component and reside at redshifts of $z<1$. The morphology can be determined from either visual inspection or meeting the sample selection requirement of a Se\'rsic index of $n<1.2$. Finally, the relations should not be applied to galaxies classified as having AGNs, as the AGNs could contaminate observations from the FUV to IR \citep{2015A&A...576A..10C}.

Additionally, the inclination estimates used in this study rely on the various assumptions made in \citet{2021ApJ...923...26D} to convert axis ratio to inclination. If alternative methods and assumptions are used, they have been shown to typically result in comparable inclination estimates. However, they tend to underestimate the uncertainty on inclination when simply propagating the axis ratio uncertainty (see Section~3 of \citealt{2021ApJ...923...26D} for details). Therefore, the relations presented in this study will be applicable even if the inclinations are estimated from an axis ratio via a different method.

\begin{figure}[t!]
\centerline{
\includegraphics[width=8.75cm]{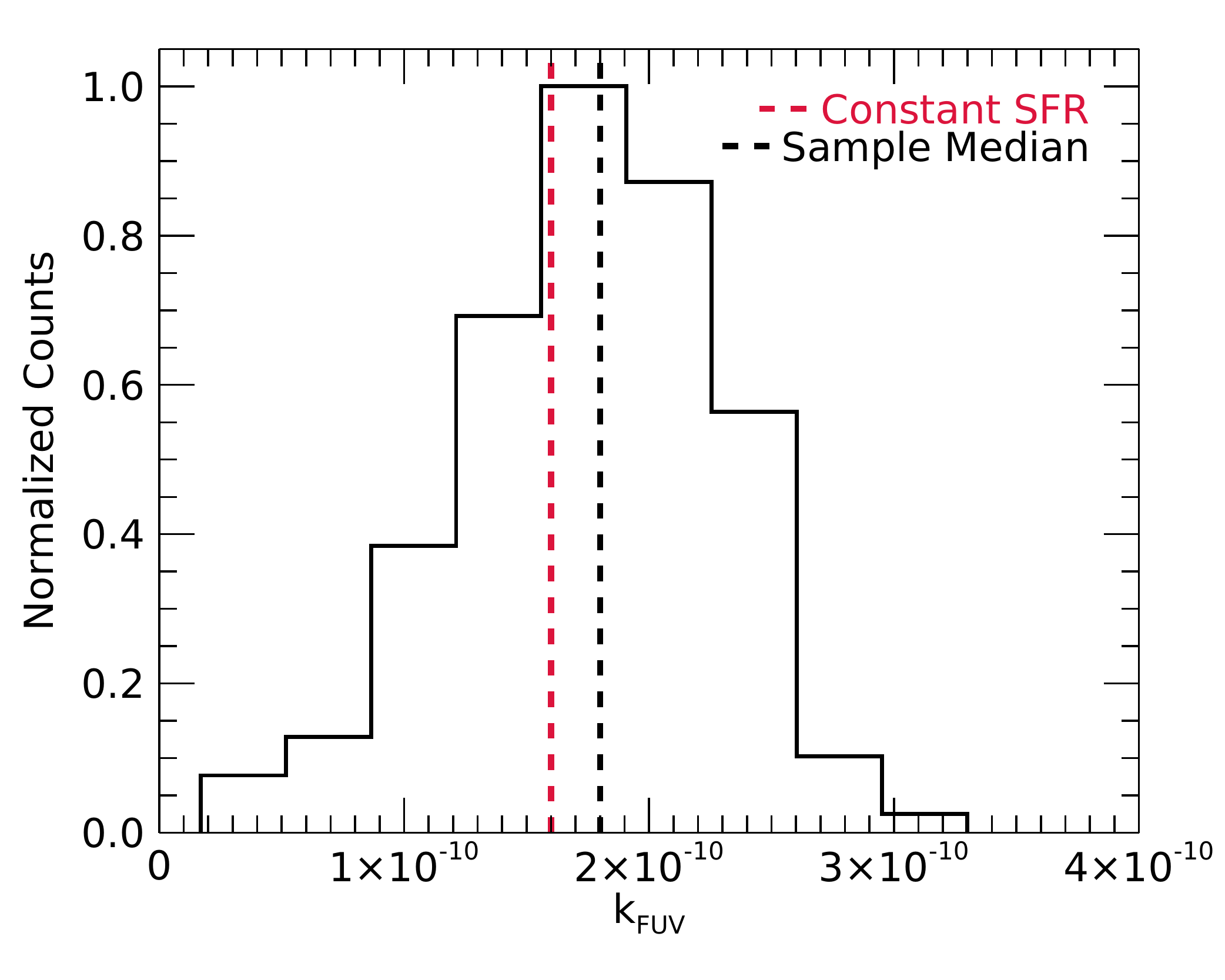}
}
\caption{
Histogram of $k_{\rm FUV}$ for the CANDELS and SINGS/KINGFISH samples. The dashed red line gives the value of $k_{\rm FUV}$ assuming a constant SFR over the last 100 Myr ($k_{\rm FUV}=1.6\times10^{-10}$), and the dashed black line gives the sample median.
}
\label{fig:kuv}
\end{figure}

While the relations presented in this paper derive an unattenuated FUV luminosity, the actual quantity of interest is the SFR. To determine the SFR from the unattenuated FUV luminosity, a conversion factor $k_{\rm UV}$ (specifically, $k_{\rm FUV}$) for use in Equation~\ref{eq:LUVtoSFR} must be selected. A variety of values can be theoretically determined depending on the assumed IMF, metallicity, and SFH, with a constant SFH over the last 100 Myr typically being assumed \citep[e.g.,][]{1998ARA&A..36..189K,2011ApJ...737...67M,2012ARA&A..50..531K}. For our assumed IMF and metallicity, this constant SFH results in $k_{\rm FUV}=1.6\times10^{-10}$. However, while the galaxies in our sample assume the same IMF and metallicity, they each have a unique SFH, which will result in each galaxy having a unique value of $k_{\rm FUV}$. In Figure~\ref{fig:kuv}, we show how these unique $k_{\rm FUV}$ values compare to the constant value of $k_{\rm FUV}$ assuming a constant SFH, which is shown as the dashed red line. On average, the galaxies in our sample have a higher $k_{\rm FUV}$ than this constant value but are consistent when considering the relatively large uncertainty with a sample median and standard deviation of $k_{\rm FUV}=(1.80\pm0.54)\times10^{-10}$. Since $k_{\rm FUV}$ is dependent on the SFH, we investigated parameterizing $k_{\rm FUV}$ as function of FUV--NIR color. However, we found that any parameterization of $k_{\rm FUV}$ with color yielded results consistent with those for a constant value of $k_{\rm FUV}$. Therefore, we recommend using the theoretical constant value of $k_{\rm FUV}=1.6\times10^{-10}$ with a propagated uncertainty of $ 0.54\times10^{-10}$ when using our relations to convert FUV luminosity to SFR.

\section{Summary} \label{sec:Summary}

We analyzed how both hybrid SFR estimators and the $A_{\rm FUV}$--$\beta$ relation depend on inclination and derived new relations to account for this inclination dependence. This analysis utilized the inclination-dependent attenuation module in the SED fitting code \texttt{Lightning}, which was applied to a sample of 133 galaxies from the CANDELS fields along with 18 local galaxies from the SINGS/KINGFISH sample in \citet{2017ApJ...837...90D}. All galaxies were selected to be disk-dominated via their Se\'rsic index and/or a visual inspection.

For the hybrid SFR estimators, we found that the UV+IR correction factor $a_{\rm corr}$ was found to be highly dependent on the inclination of a galaxy in addition to its sSFR. Since the sSFR is not an observable quantity, a {rest-frame FUV--NIR color was used as a proxy along with inclination to derive the parametric relation for $a_{\rm corr}$ given in Equation~\ref{eq:acorrcolorinc}. The relation was a simple linear fit of FUV--NIR color to $a_{\rm corr}$, with the linear coefficients being polynomials of inclination. These polynomial coefficients were presented in Table~\ref{table:colorcoeff} for four different FUV--NIR colors. These relations were shown to predict values of $a_{\rm corr}$ that were highly consistent with the data and properly account for any inclination dependence.

As for the $A_{\rm FUV}$--$\beta$ relation, we derived two different sets of $\beta$ to account for the potential contamination of observations by the rest-frame UV bump feature. The first set includes the rest-frame GALEX FUV and NUV bandpasses, with the NUV bandpass overlapping with the UV bump. The second set includes the rest-frame GALEX FUV and HST WFC3/F275W bandpasses, both of which avoid the bump feature. For both sets of $\beta$, we found that there is a definite inclination dependence with edge-on galaxies having a higher $A_{\rm FUV}$ by 1-2 mag for a given value of $\beta$ compared to more face-on galaxies. To derive our inclination-dependent $A_{\rm FUV}$--$\beta$ relation for each set, we fit the relation given in Equation~\ref{eq:AUVbeta} to our data. These fits resulted in the expected trends of an increase in $a_\beta$ and a constant $\beta_0$ with inclination for $1-\cos i \leq 0.75$. However, at higher inclinations, $a_\beta$ and $\beta_0$ deviated from these expected trends, with both decreasing with increasing inclination. We attributed these deviations to various simplifying assumptions within the $A_{\rm FUV}$--$\beta$ relation. Regardless, we fitted polynomials for the full range of inclination to $a_\beta$ and $\beta_0$, whose coefficients were presented in Table~\ref{table:abetacoeff}, and noted that the linearity of the $A_{\rm FUV}$--$\beta$ relation is likely too simplified for highly inclined galaxies.

The results of this work illustrate that inclination can significantly affect the derived SFR in disk-dominated galaxies when using UV SFR tracers. We find that including an inclination dependence in these tracers is critical for more accurate SFR estimates. In future work, we plan to apply the inclination-dependent attenuation module in \texttt{Lightning} to a more complete sample of galaxies that have sizable bulge components, rather than a purely disk-dominated sample. We intend to see how the bulge component of a galaxy affects the inclination dependence of our results and check if similar relations apply to the broader disk galaxy population.

\begin{acknowledgements}
We acknowledge and thank the anonymous referee for the valuable and insightful comments that helped improve the quality of this paper. We gratefully acknowledge support from NASA Astrophysics Data Analysis Program (ADAP) grant 80NSSC20K0444 (K.D., R.T.E., B.D.L., E.B.M.) and NASA award No. 80GSFC21M0002 (A.B.). K.G. was supported by an appointment to the NASA Postdoctoral Program at Goddard Space Flight Center, administered by Oak Ridge Associated Universities under contract with NASA. This work is based on observations taken by the CANDELS Multi-Cycle Treasury Program with the NASA/ESA HST, which is operated by the Association of Universities for Research in Astronomy, Inc., under NASA contract NAS5-26555. This work has made use of the NASA/IPAC Extragalactic Database (NED), which is funded by the National Aeronautics and Space Administration and operated by the California Institute of Technology, and the Arkansas High Performance Computing Center, which is funded through multiple National Science Foundation grants and the Arkansas Economic Development Commission. We acknowledge the usage of the HyperLeda database (\url{http://leda.univ-lyon1.fr}).
\end{acknowledgements}

\facilities{HST, Spitzer, Herschel, Blanco, CFHT, ESO:VISTA, LBT, Mayall, Subaru, UKIRT, VLT:Melipal, VLT:Yepun}

\software{\texttt{Lightning} (Eufrasio et al. 2017; Doore et al. 2021)}

\bibliographystyle{aasjournal}
\bibliography{IncDepSFR}

\appendix

\section{Spectroscopic Redshift Catalog} \label{sec:RedCatalog}

The spectroscopic redshifts assigned to sources in the CANDELS fields were compiled from various sources. For the GOODS-N, we used the relatively comprehensive CANDELS redshift catalog from \citet{2019ApJS..243...22B}. For the GOODS-S, we compiled spectroscopic redshifts from the Chandra Deep Field-South ``master spectroscopic catalog,''\footnote{\url{https://www.eso.org/sci/activities/garching/projects/goods/MasterSpectroscopy.html}} ACES \citep{2012MNRAS.425.2116C}, and VANDELS spectroscopic surveys \citep{2021A&A...647A.150G} that were not already included in the GOODS-S CANDELS redshift and mass catalog \citep{2015ApJ...801...97S}. These sources were then cross-matched to the nearest CANDELS source within $0.5^{\prime\prime}$. If a source in the master catalog, ACES, or VANDELS had a higher reliability flag than what was in the CANDELS catalog, we replaced the CANDELS spectroscopic redshift with the more reliable measurement. For the EGS, we cross-matched spectroscopic redshift sources from the DEEP2+3 survey data release 4 \citep{2004ApJ...609..525C,2006AJ....132.2159W,2011ApJS..193...14C,2012MNRAS.419.3018C,2013ApJS..208....5N}\footnote{\url{https://deep.ps.uci.edu}} to the nearest source within $0.5^{\prime\prime}$ in the CANDELS EGS multiband catalog. For the COSMOS field, we cross-matched spectroscopic redshift sources from IMACS \citep{2009ApJ...696.1195T}, zCOSMOS data release 3 \citep[DR3;][]{2009ApJS..184..218L}\footnote{\url{https://www.eso.org/qi/catalog/show/65}}, FMOS \citep{2015ApJS..220...12S}, LEGA-C DR3 \citep{2016ApJS..223...29V}\footnote{\url{https://www.eso.org/qi/catalog/show/379}}, hCOSMOS \citep{2018ApJS..234...21D}, DEIMOS \citep{2018ApJ...858...77H}, and C3R2 \citep{2019ApJ...877...81M} to the nearest source within $0.5^{\prime\prime}$ in the CANDELS COSMOS multiband catalog. If a galaxy had redshifts from multiple surveys, the most reliable redshift was used. For the UDS field, we included any spectroscopic redshifts from the UDSz spectroscopic catalog \citep{2013MNRAS.433..194B,2013MNRAS.428.1088M}\footnote{\url{https://www.nottingham.ac.uk/astronomy/UDS/UDSz/}}, VANDELS spectroscopic survey, and C3R2 that were not already included in the UDS CANDELS redshift and mass catalog \citep{2015ApJ...801...97S} by cross-matching them to the nearest source within $0.5^{\prime\prime}$. If a source in UDSz, VANDELS, or C3R2 had a higher reliability flag than what was in the CANDELS catalog, we replaced the CANDELS spectroscopic redshift with the more reliable measurement.

\end{document}